\documentclass[16pt, doublecolumn]{IEEEtran}
\IEEEoverridecommandlockouts
\usepackage{pdfcomment}
\usepackage{cite}
\usepackage{amsthm}
\usepackage{amsmath,amssymb,amsfonts}
\usepackage{graphicx}
\usepackage{textcomp}
\usepackage{xcolor}
\usepackage{amsthm,amsmath,amssymb}
\usepackage{mathrsfs}
\usepackage{amsmath,bm}
\usepackage{algorithm} 
\usepackage{mathtools}
\usepackage{epsf}
\usepackage{cleveref}
\usepackage{verbatim}
\usepackage{algorithmic}
\usepackage{multirow}
\usepackage{float}  %设置图片浮动位置的宏包
\usepackage{subfigure}  %插入多图时用子图显示的宏包
\usepackage{colortbl}
\usepackage{pifont}

\def\BibTeX{{\rm B\kern-.05em{\sc i\kern-.025em b}\kern-.08em
		T\kern-.1667em\lower.7ex\hbox{E}\kern-.125emX}}
\begin{document}
	
\title{Optimization for Master-UAV-powered Auxiliary-Aerial-IRS-assisted IoT Networks: An Option-based Multi-agent Hierarchical Deep Reinforcement Learning Approach  }

\author{Jingren~Xu,~Xin~Kang,~\IEEEmembership{Senior~Member,~IEEE},~Ronghaixiang~Zhang,\\~Ying-Chang Liang,~\IEEEmembership{Fellow,~IEEE}, and Sumei Sun,~\IEEEmembership{Fellow,~IEEE}
	
	\thanks{This work was supported in part by the National Natural Science Foundation of China under Grants U1801261, by the National Key R\&D Program of China under Grant 2018YFB1801105, by the Key Areas of Research and Development Program of Guangdong Province, China, under Grant 2018B010114001, by the Science and Technology Development Fund, Macau SAR, under Grant 0009/2020/A1, by the Central Universities under Grant ZYGX2019Z022 and by the 111 Project under Grant B20064. Part of this work was presented in IEEE Global Communications Conference 2021 \cite{globecom2021_xu}.}
	\thanks{J. Xu, X. Kang, R. Zhang and Y.-C. Liang are with the Center for Intelligent Networking and Communications (CINC), University of Electronic Science and Technology of China (UESTC), Chengdu 611731, China (e-mail: {jingrenxu@163.com; kangxin83@gmail.com; fredrickzhang@yahoo.com; liangyc@ieee.org}).}
	\thanks{S. Sun is with the Institute for Infocomm Research, Agency for
		Science, Technology and Research, Singapore 138632 (e-mail: {sunsm@i2r.a-star.edu.sg}).}
}
\maketitle
\begin{abstract}
	This paper investigates a master unmanned aerial vehicle (MUAV)-powered Internet of Things (IoT) network, in which we propose using a rechargeable auxiliary UAV (AUAV) equipped with an intelligent reflecting surface (IRS) to enhance the communication signals from the MUAV and also leverage the MUAV as a recharging power source. Under the proposed model, we investigate the optimal collaboration strategy of these  energy-limited UAVs to maximize the accumulated throughput of the IoT network. Depending on whether there is charging between the two UAVs, two optimization problems are formulated. To solve them, two multi-agent deep reinforcement learning (DRL) approaches are proposed, which are centralized training multi-agent deep deterministic policy gradient (CT-MADDPG) and multi-agent deep deterministic
	policy option critic (MADDPOC).  It is shown that the CT-MADDPG can greatly reduce the requirement on the computing capability of the UAV hardware, and the proposed MADDPOC is able to support low-level multi-agent cooperative learning in the continuous action domains, which has great advantages over the existing option-based hierarchical DRL that only support single-agent learning and discrete actions.
\end{abstract}
\begin{IEEEkeywords}
	UAV, IRS, rechargeable, multi-agent deep reinforcement learning, hierarchical
\end{IEEEkeywords}

\section{Introduction}
The Internet of Things (IoT), aiming at providing a pervasive connection of digital devices, is a novel cutting edge technology for the future digital world \cite{8030479}, \cite{9499093}. It has affected various aspects of our life and supports the development of smart city, smart agriculture, smart farming, smart manufacturing, and etc \cite{Mohanty20162162}. With diverse use cases anticipated in IoT, the types of IoT devices are expected to be diversified, and the deployment of different IoT devices are also expected to vary significantly. Some types of IoT devices, such as sensors, are expected to be  deployed sparsely in certain scenarios such as smart agriculture. For these scenarios, the traditional base station (BS) based communication solutions may not be cost-effective or high-efficient due to the long-distance path loss. 

In contrast to the traditional terrestrial BS, unmanned aerial vehicle (UAV) communications not only benefit from the considerable channel gain brought by higher probability of line-of-sight (LoS) channel, but also enjoy the coverage gain by taking advantage of high maneuverability \cite{7470933}. Thus,  UAVs have recently drawn significant attention in data collection tasks of IoT devices \cite{8255734}. Related optimization problems of deployment location, trajectory design and resource allocation in UAV based wireless networks have been studied in [7]-[10]. Yang $et~al.$ \cite{9222571} maximized the energy efficiency in a UAV-assisted backscatter communication network by jointly optimizing the devices' wake-up schedule, power control, and the UAV’s trajectory. Ye $et~al.$ \cite{9080561} proposed a dynamic time division multiple access (TDMA) structure for full-duplex UAV-enabled wireless powered network. 
In the multi-UAV cooperative scenario, \cite{8247211} proposed an iterative algorithm based on block coordinate descent to solve mixed integer non-convex optimization problem, where users' communication
scheduling, UAVs’ trajectory, and power control were highly coupled. In 3-D space, a novel framework for full-duplex orthogonal frequency-division multiple access UAV-enabled wireless-powered IoT networks was proposed in \cite{ye2021Joint}, 
where a swarm of UAVs was deployed to simultaneously
charge all devices and then fly to new locations to collect
information from scheduled devices. However, limited battery capacity on board is a key factor affecting  the performance of UAVs in practice, which was ignored in aforementioned work. Although \cite{9222571} took it into account, there's no guarantee of successful return after task termination. Considering the downsides of traditional charging methods, e.g. suspending tasks or wasting time in landing, there are some works dedicated to prolonging the lifetime of UAV networks. In \cite{9177252}, sustainable solar-powered UAVs were applied in random access IoT networks, where UAVs can be recharged through solar radiation. This method is highly weather-dependent and cannot be used in bad weather areas. Another promising technique is wireless-powered UAV \cite{rechargeable_UAV}. Such researches usually deploy static charger with high-voltage power lines on the ground \cite{2020Design, 9488324}. However, the bad channel conditions requires a demanding charging distance. As a result, UAV has to find a trade-off between charging and communications.

On the other hand, the intelligent reflecting surface (IRS), benefiting from the breakthrough on the fabrication of inexpensive programmable meta-material, can improve the channel environment and extend the coverage \cite{9326394}. Existing researches on IRS-assisted UAV communications can mainly be categorized into two groups: 1) IRS is deployed at a terrestrial point, e.g. the surface of buildings, to enhance the signals from the UAV \cite{9145224}, \cite{9293155}. In \cite{9416239}, the average worst-case secrecy rate was maximized in an IRS-aided UAV secure communication system. Under the circumstance, the IRS is supposed to be in sight of both the UAV and the users \cite{9090356}, which is highly dependent on the deployment environment. 2) IRS is deployed in the air to achieve 360° panoramic full-angle reflection and enjoy the LoS air-to-ground (A2G) channel\cite{9145305}. In \cite{9234511} and \cite{9130430}, an integrated UAV-IRS was implemented as a relay node to provide an enhanced link to the ground users. The beamforming as well as the UAV's position was optimized to achieve a better channel quality. Besides, the authors in \cite{9539168} considered an IRS-assisted multi-layer UAVs network, where a quasi-stationary UAV equipped with an IRS was placed at higher altitudes to serve smaller UAVs below. Howerever, the works aforementioned treat  UAVs as spot-deployable carriers and ignore the gains brought by their high mobility.

Deep reinforcement learning (DRL) methods can deal with non-convex optimization problems with high complexity and is widely used in various UAV-assisted communication scenarios \cite{9043893}. \cite{9248522} proposed a deep Q-learning based method for UAV trajectories design and backscatter device scheduling. One single UAV is inefficient especially when the ground devices are widely distributed. Thus in \cite{8676325}, a fully-distributed control solution based on the deep deterministic policy gradient (DDPG) algorithm \cite{DDPG} was designed to navigate a UAV swarm for fair communication coverage. However, the aforementioned works ignored the limited on-board battery capacity which makes it unpractical for time-consuming tasks \cite{8933037}. This constraint was considered in \cite{9171468} and a distance-related reward was designed to help the UAV learn to return to the charging station. However, the method used in \cite{9171468}, i.e., overlaying a smaller reward with a larger one, leads to training instability. The authors in \cite{9200357} proposed an option-based DRL methods with hierarchical control to deal with such problems. To be specific, each UAV makes a choice among options, such as communication, charging and task termination, according to upper-level policies,  and execute specific actions based on low-level policies. But the low-level policies in \cite{9200357}, e.g. the trajectories of UAVs, were replaced by the artificial strategies for simplicity without optimization. In addition, fully-distributed training costs more time or needs more hardware resources on parameters optimization compared with the centralized method, which leads to unsatisfactory  performance in certain cases \cite{MAPPO}. 

To deal with the limitations of aforementioned works, in this paper, we propose using a rechargeable Auxiliary UAV (AUAV) equipped with an IRS to assist the data collection of  Master UAV (MUAV)-powered IoT network.  To be specific, the IRS-equipped rechargeable AUAV is used to enhance the communication signals from the MUAV, and it also leverages the MUAV as a flying recharging power source. 

The main contributions are listed as follows:
\begin{itemize}
	\item[$\bullet$] We propose a novel master-UAV-powered auxiliary-aerial-IRS-assisted IoT network, which is different from existing works as
	follows: a) Both the MUAV and the AUAV are no longer deployed at fixed  points. Instead, both are allowed to move freely within the IoT network. In this way, our model can  enjoy the gains brought by both UAVs' high mobility and the IRS's signal enhancement. b) A wireless charging mechanism is added, through which the AUAV can harvest energy from the MUAV during the task  instead of suspending task to recharge.
	\item[$\bullet$] Under the proposed new model, two network throughput maximization problems are formulated, depending on whether there is wireless charging between the AUAV and the MUAV. We investigate the optimal collaboration strategies (including power allocation, UAV trajectories, task termination, and charging) of these UAVs to maximize the accumulated throughput of the IoT network. Especially, unlike existing works, we assume both the AUAV and the MUAV are battery-limited, and we request both of them to return the originating point when the data collection task terminates,  which makes our optimization problem more practical but more challenging.  
	\item[$\bullet$] To solve the formulated optimization problems, two new multi-agent DRL approaches are proposed: a) For the problem without charging between UAVs, we propose a centralized training multi-agent DRL approach, referred to as centralized training multi-agent deep deterministic policy gradient (CT-MADDPG), to deal with the learning of low-level policies. Unlike the
	traditional MADDPG, the CT-MADDPG removes many unnecessary parameters, which greatly reduces the requirement on the computing hardware.  b). For the problem with charging between UAVs, we model the optimization problem as a multi-agent semi-markov decision process (SMDP), and we propose a hybrid hierarchical DRL approach , which is named as multi-agent deep deterministic policy option critic (MADDPOC). Unlike the existing 
	option-based hierarchical DRL that only support single-agent learning and discrete
	actions, the proposed MADDPOC not only supports low-level multi-agent cooperative learning but also support  continuous actions.
\end{itemize}

The remainder of this paper is organized as follows. Section~\ref{system_model} describes the system model. In Section \ref{task_description_p_form}, the task is elaborated and the problems are formulated. Section \ref{madrl} shows the CT-MADDPG algorithm for the non-charging scenario, while in Section \ref{mahdrl}, we propose the MADDPOC algorithm as a hierarchical extension of the CT-MADDPG for the charging scenario. In Section, \ref{result} we present simulation results for the proposed algorithms. Finally,
we conclude the paper in Section \ref{conclusion}.

\textit{Notations:} A scalar $x$, a vector $\mathbf{x}$, and a matrix $\mathbf{X}$ are given by $x$, $\mathbf{x}$, and $\mathbf{X}$, respectively. $\mathbb{R}^{M \times N}$ and $\mathbb{C}^{M \times N}$ represent the space of a $M \times N$ matrix with real and complex entries, respectively. $\left\| \cdot \right\|$ refers to a vector norm. $\left (\cdot \right )^{\mathrm{H}}$ denotes the conjugate transpose of a matrix. $\otimes$ and $\odot$ donate the Kronecker product and Hadamard product of two matrices , respectively.
$\mathcal{C}\mathcal{N} \left( 0, \sigma^2\right) $ represents a circularly symmetric complex Gaussian
(CSCG) vector’s distribution with zero mean and noise variance $\sigma^2$, and $\sim$ means “distributed as”. $diag \left( x_1,\cdots, x_k\right) $ represents a diagonal matrix with the elements given by $\left\lbrace x_1,\cdots, x_k\right\rbrace $. The operator$\mod$ returns the
remainder after division.

\section{System Model}\label{system_model}
\begin{figure}[t]
	\centering
	\includegraphics[height=.75\columnwidth, width=.85\columnwidth] {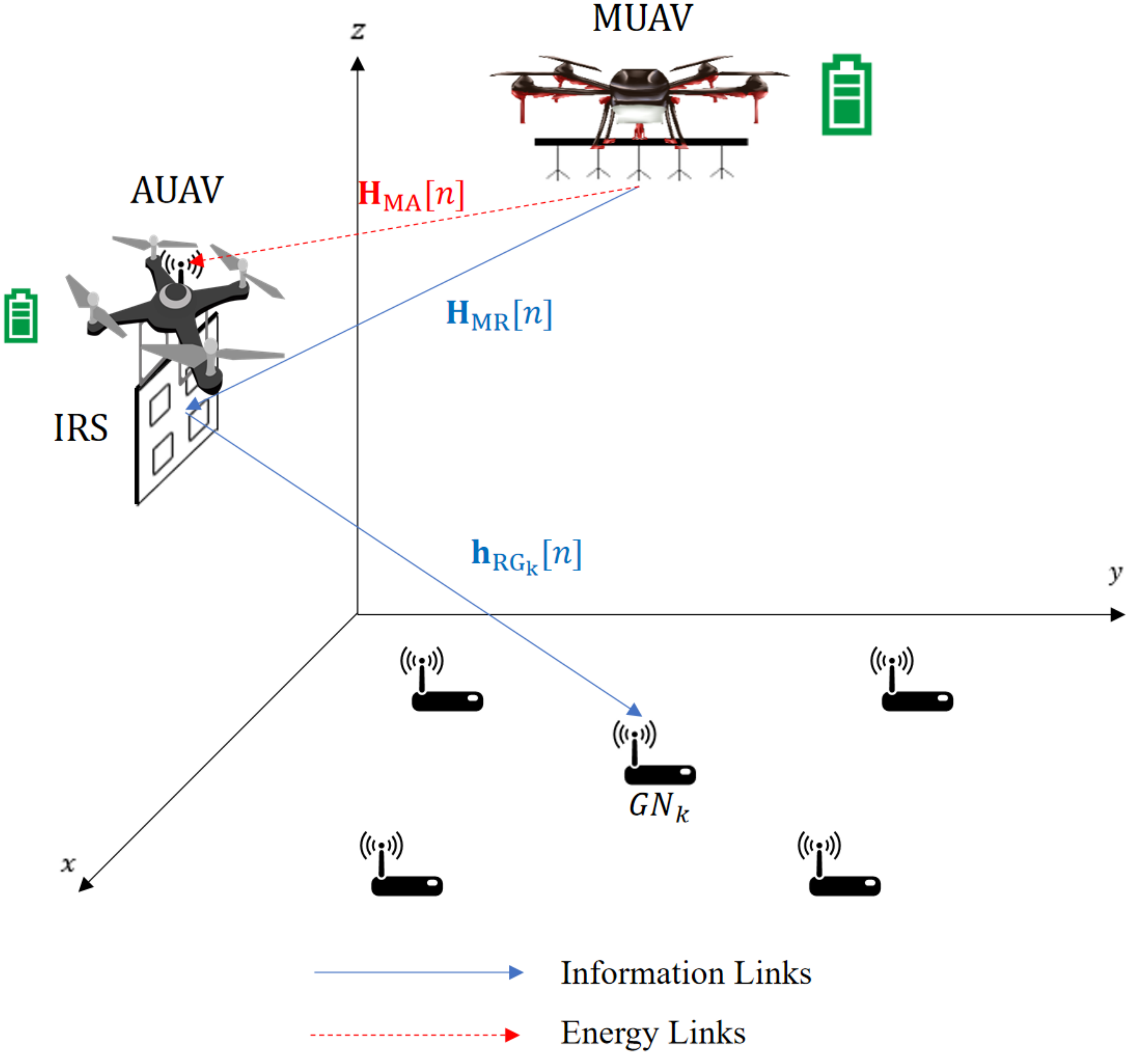}
	\caption{System model.} 
	\label{fig:scenario}
\end{figure}

We consider a dual-UAV-enabled downlink TDMA IoT network serving $K$ single-antenna GNs with fixed locations assisted by an aerial IRS as shown in Fig.~\ref{fig:scenario}. The MUAV carries an uniform plane array (UPA) with $S_{\mathrm{M}}= S_{\mathrm{M}_x}\times S_{\mathrm{M}_y}$ antennas for both information and wireless power transmission. Meanwhile, the AUAV is equipped with an IRS which consists of $S_{\mathrm{R}} = S_{\mathrm{R}_x}\times S_{\mathrm{R}_y}$ passive reflecting units for signal enhancement, as well as an UPA with $S_{\mathrm{A}}= S_{\mathrm{A}_x}\times S_{\mathrm{A}_y}$ antennas deployed on the top for energy harvesting. The orientation of the IRS can be adjusted flexibly to make sure it's always perpendicular to the $y$ axis. Assumed that all antennas and IRS units share a single ratio frequency (RF) chain, but each of them has an individual phase shift. In addition, a high-capacity battery with energy $E_{\mathrm{Mmax}}$ is provided to the MUAV, making it a small mobile charger, while the AUAV is equipped with a lightweight battery with low capacity $E_{\mathrm{Amax}}$ in consideration of the on-board IRS's weight and UAV's restricted loading capacity, and needs to be charged wirelessly by the MUAV. The communication and charging processes are assumed to be simplex, i.e., only one is allowed per slot, and the request for charging is available at any time \cite{8424210}. Besides, two UAVs operate at the fixed altitudes of $H_{\mathrm{M}},H_{\mathrm{A}}$, where $H_{\mathrm{M}}>H_{\mathrm{A}}$, respectively. The total serving time $T$ is divided into $N_{\tau}$ equal-length slots with duration time $\delta_{\tau}$ (s), i.e., $T=N_{\tau}\delta_{\tau}$, where the value of $N_{\tau}$ depends on the specific policy. Given the communication fairness, the MUAV will automatically change its linked GN per $N_{u}$ slots starting with the GN closest to the initial UAV charging station, i.e. $k =(k+\lfloor\frac{n+1}{N_u}\rfloor) \mod (K+1)$, where  $k \in \left\lbrace 1,\cdots, K\right\rbrace $ labels the index of currently linked GN.

For the sake of convenience, we donate the time-varying horizontal locations $\textbf{l}_\mathrm{M}\left [ n \right ]=\left [ x_{\mathrm{M}}\left[n \right], y_{\mathrm{M}}\left[n \right]  \right ], \textbf{l}_\mathrm{A}\left [ n \right ]=\left [ x_{\mathrm{A}}\left[n \right] y_{\mathrm{A}}\left[n \right]  \right ], n\in\left\lbrace 0,\cdots,N_\tau \right\rbrace ,$ for the MUAV and AUAV, respectively, and $\textbf{l}_{\mathrm{G}_{k}}=\left[x_{\mathrm{G}_{k}},y_{\mathrm{G}_{k}}  \right], \textbf{l}_\mathrm{C}=\left [ x_{\mathrm{C}}, y_{\mathrm{C}} \right ]$ for the stationary GN $k$ and charging station. The distance between the MUAV and AUAV, AUAV and GN $k$, MUAV and charging station, AUAV and charging station, are donated by $d_\mathrm{MA}\left [ n \right ], d_\mathrm{AG_k}\left [ n \right ], d_\mathrm{MC}\left [ n \right ]$ and $d_\mathrm{AC}\left [ n \right ]$, respectively. 

\begin{figure*}[t!]
	\subfigure[]
	{
		\begin{minipage}{0.32\linewidth}
			\centering
			\includegraphics[width=2.3in]{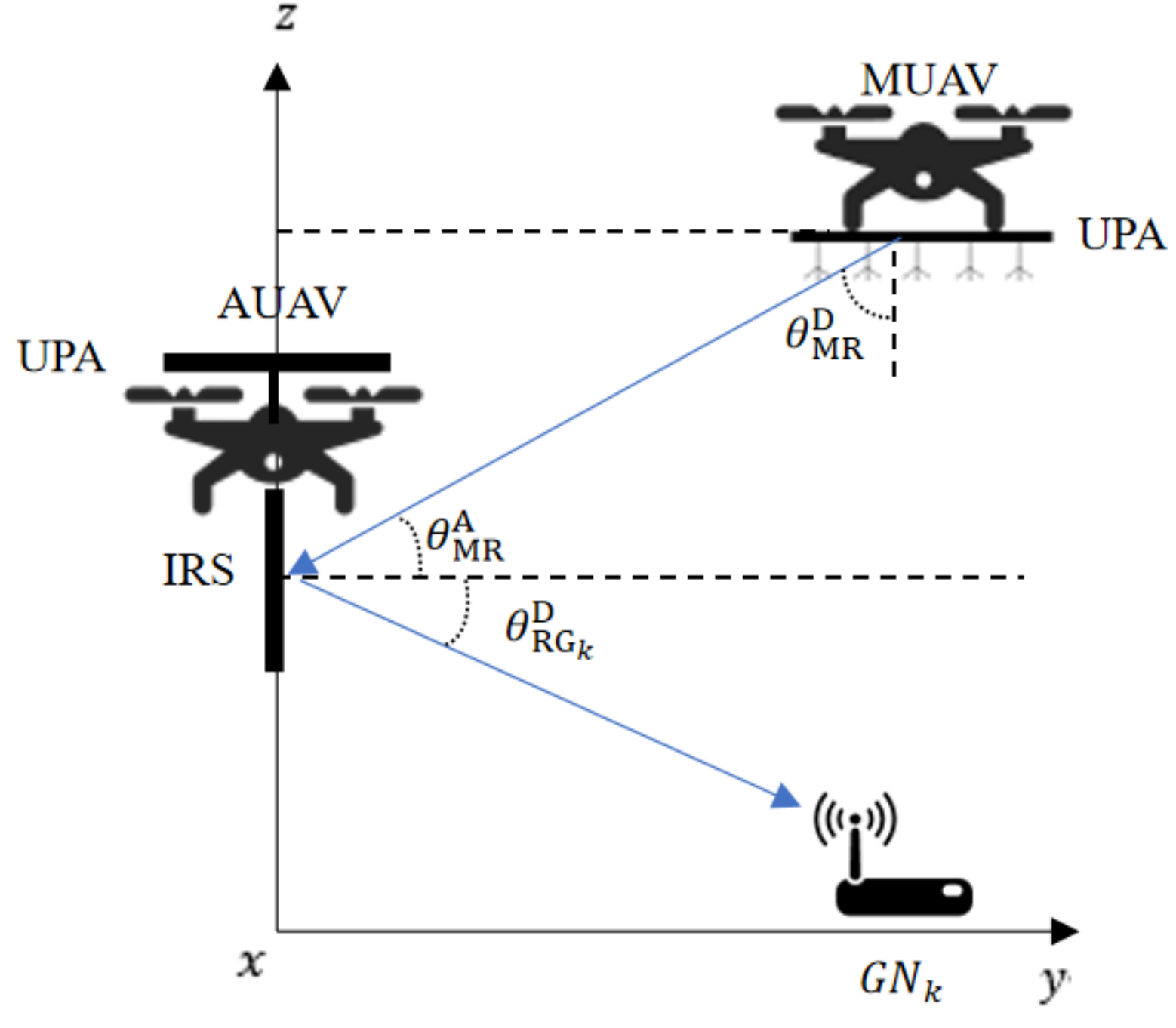}
			\label{fig:AoA1} 
		\end{minipage}
	}
	\subfigure[]
	{
		\begin{minipage}{0.32\linewidth}
			\centering			\includegraphics[width=2.3in]{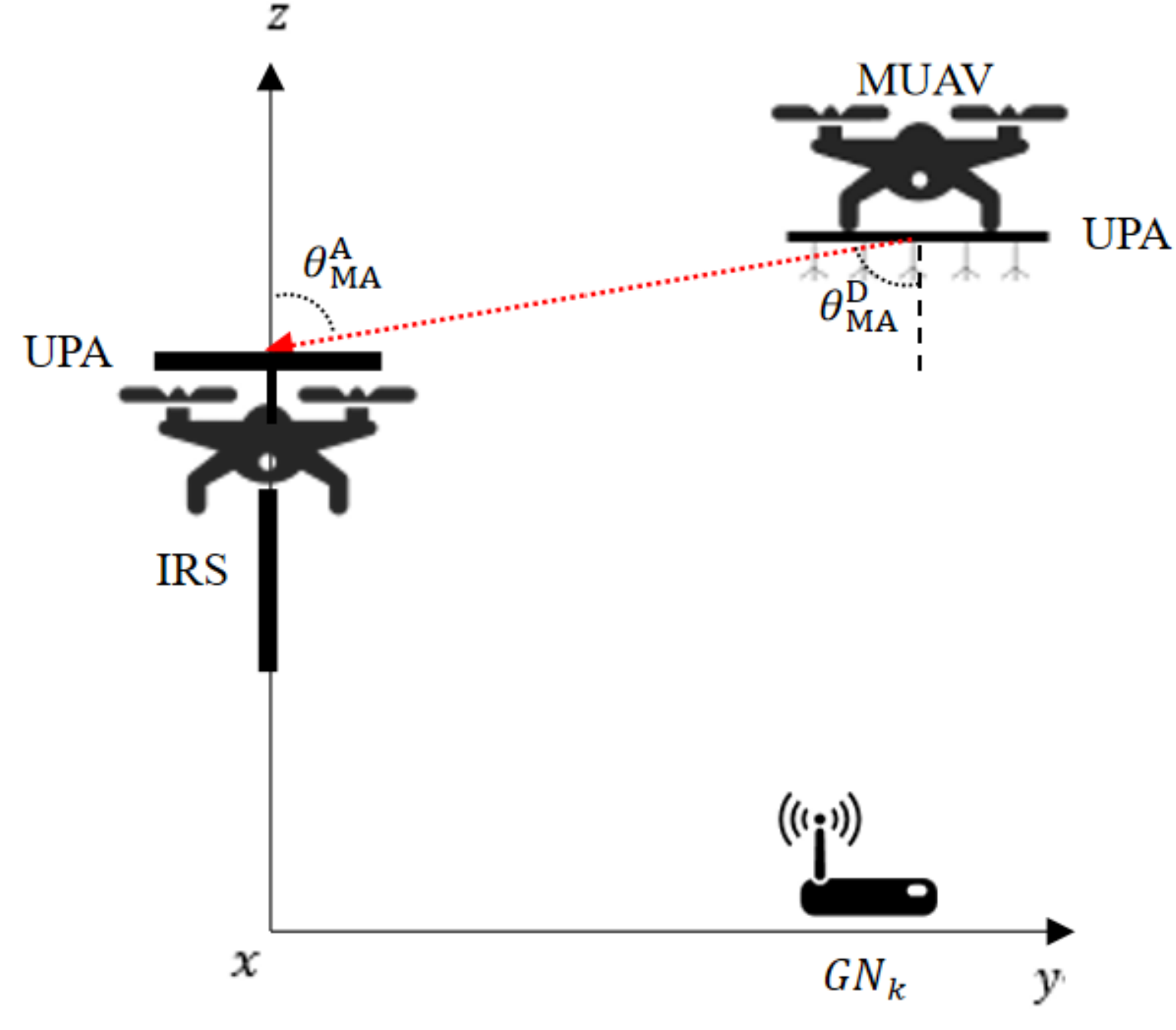}
			\label{fig:AoA2}
		\end{minipage}
	}
	\subfigure[]
	{
		\begin{minipage}{0.32\linewidth}
			\centering			\includegraphics[width=2.3in]{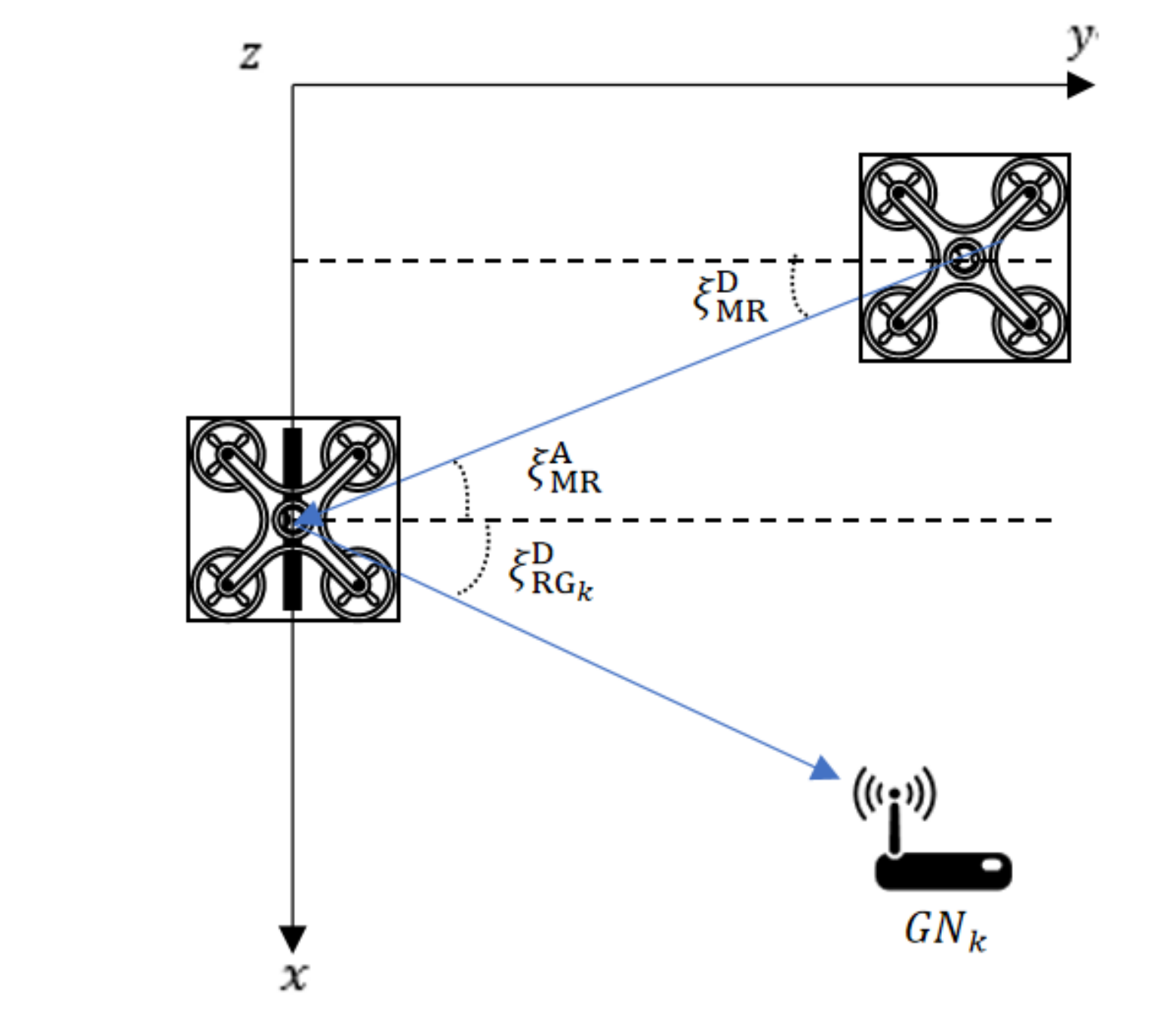}
			\label{fig:AoA3}
		\end{minipage}
	}
	
	\caption{(a) Vertical AoAs/AoDs between the MUAV, IRS and GN $k$. (b) Vertical AoAs/AoDs between the MUAV and AUAV's UPA when charging. (c) Horizontal AoAs/AoDs between the MUAV, AUAV and GN $k$.}
	\label{fig:AoA/AoD}
\end{figure*}

\subsection{Channel Model}
The transmission links between the MUAV, AUAV and GNs can be regarded as A2G channels which suffer from the extra path loss brought by the buildings, trees and other obstacles with high altitude. Such channel is generally donated by a probabilistic-average-based LoS (PLoS) model, which is expressed as 
\begin{align}
	\overline{PL}(d,\Delta H)= PL^\mathrm{FS}(d) +\Gamma^\mathrm{LoS}(d,\Delta H) \eta_\mathrm{LoS} \nonumber \\ +  (1-\Gamma^\mathrm{LoS}(d,\Delta H)) \eta_\mathrm{NLoS},\label{eq:1}
\end{align}
where $ PL^\mathrm{FS}\left(d\right) = 20\log \frac{4\pi f_c d}{c}$ is the free space path loss related to the distance $d$, $f_c$ donates the carrier frequency, $c$ refers to the velocity of light, and  $\eta_\mathrm{LoS}, \eta_\mathrm{NLoS}$ are the excessive path loss component for LoS and NLoS links, respectively. Besides, $\Gamma^\mathrm{LoS}$ donates the probability of
LoS connectivity, which is modified by a simple sigmoid function \cite{6863654}, given by
\begin{align}
	\Gamma^\mathrm{LoS}(d,\Delta H) =  \frac{1}{1+\eta_a \exp \left(-\eta_b\left(\arcsin \left( \frac{\Delta H}{d}\right)  - \eta_a  \right)  \right) }  \label{eq:3},
\end{align}
where $\Delta H$ refers to the altitude difference, $\eta_a$ and $\eta_b$ are constants related to the type of propagation environment.

Therefore, the channel response vectors from MUAV to IRS, from MUAV to AUAV's UPA and from IRS to GN $k$ at time slot $n$, donated by $\mathbf{H}_\mathrm{MR}\left [ n \right ]\in \mathbb{C}^{S_\mathrm{R} \times S_\mathrm{M}},\mathbf{H}_\mathrm{MA}\left [ n \right ]\in \mathbb{C}^{S_\mathrm{A} \times S_\mathrm{M}}$ and $\mathbf{h}_\mathrm{RG_k}\left [ n \right ]\in \mathbb{C}^{S_\mathrm{R}\times 1},$ respectively, can be modeled as 
\begin{align}
	\mathbf{H}_\mathrm{MR}\left [ n \right ] = \sqrt{\alpha_{\rm{MR}}[n]}\mathbf{a}\left(\theta_{\rm{MR}}^{\rm{A}}\left [ n \right ], \xi_{\rm{MR}}^{\rm{A}}\left [ n \right ], S_{{\rm{R}}_x}, S_{{\rm{R}}_y} \right) \nonumber \\ \otimes \mathbf{a}\left(\frac{\pi}{2}-\theta_{\rm{MR}}^{\rm{D}}\left [ n \right ], \xi_{\rm{MR}}^{\rm{D}}\left [ n \right ], S_{\rm{M}_x},S_{\rm{M}_y} \right) ^ {H},\label{eq:H_MR}\\ 
	\mathbf{H}_\mathrm{MA}\left [ n \right ] =\sqrt{\alpha_{\rm{MA}}[n]} \mathbf{a}\left(\theta_{\rm{MA}}^{\rm{A}}\left [ n \right ], \xi_{\rm{MR}}^{\rm{A}}\left [ n \right ], S_{{\rm{A}}_x}, S_{{\rm{A}}_y} \right) \nonumber \\ \otimes \mathbf{a}\left(\theta_{\rm{MA}}^{\rm{D}}\left [ n \right ], \xi_{\rm{MR}}^{\rm{D}}\left [ n \right ], S_{\rm{M}_x},S_{\rm{M}_y} \right) ^ {H},\label{eq:H_MA} \rm{and}\\
	\mathbf{h}_\mathrm{RG_k}\left [ n \right ] = \sqrt{\alpha_{\rm{RG_k}}[n]} \mathbf{a}\left(\theta_\mathrm{RG_k}^{\rm{D}}\left [ n \right ], \xi_\mathrm{RG_k}^{\rm{D}}\left [ n \right ], S_{{\rm{R}}_x}, S_{{\rm{R}}_y} \right) \label{eq:h_RGk}, 
\end{align}
where $\alpha_{\rm{MR}}[n] = \alpha_{\rm{MA}}[n] = \frac{1}{\overline{PL}(d_\mathrm{MA}\left [ n \right ], H_M- H_A)}$, $ \alpha_{\rm{RG_k}}[n]=\\\frac{1}{\overline{PL}(d_\mathrm{AG_k}\left [ n \right ], H_A)}$ are the corresponding path loss defined by the PLoS model mentioned above. Besides, $\theta_{\rm{MR}}^{\rm{A}}\left [ n \right ]$, $\xi_{\rm{MR}}^{\rm{A}}\left [ n \right ]$, $\theta_{\rm{MR}}^{\rm{D}}\left [ n \right ]$, $\xi_{\rm{MR}}^{\rm{D}}\left [ n \right]$,
$\theta_{\rm{MA}}^{\rm{A}}\left [ n \right ]$, $\theta_{\rm{MA}}^{\rm{D}}\left [ n \right ]$, 
$\theta_\mathrm{RG_k}^{\rm{D}}\left [ n \right ]$ and  $\xi_\mathrm{RG_k}^{\rm{D}}\left [ n \right ]$ are the related vertical and horizontal AOAs/AODs shown in Fig~\ref{fig:AoA/AoD}. In addition, the steering vector $\mathbf{a}$ is expressed as:
\begin{align}
	\mathbf{a}\left(\theta, \xi, K_x, K_y\right)  = f_{K_x}(\sin \theta cos \xi) \otimes  f_{K_y}(\sin \theta sin \xi),
\end{align}
where $f_{M}(u) =\left[
1,e^{-j\frac{2\pi \bigtriangleup d}{\lambda_{c}}u},\cdots,e^{-j\frac{2\pi \bigtriangleup d}{\lambda_{c}}(M-1)u}\right]^{H}$, $\lambda_{c}$ denotes the carrier wavelength, and $\bigtriangleup d$ denotes the antenna or IRS unit spacing. For simplicity, we set $\Delta d/\lambda_{c}=1/2$.

\subsection{Information Transmission Model}
The direct transmission path from the MUAV to GN is ignored due to the high operation altitude. Thus, the end-to-end channel form the MUAV to GN $k$  via the IRS carried by the AUAV at time slot $n$ can be calculated as
\begin{align}
	\mathbf{h}_{\mathrm{MG}_k}\left [ n \right ] ^{H} =\mathbf{h}_{\mathrm{RG}_k}\left [ n \right ] ^{H}\mathbf{\Phi}\left[n \right]\mathbf{H}_\mathrm{MR}\left [ n \right ],\label{eq:end2endchannel}
\end{align}
where $\mathbf{\Phi}\left[n \right]=\rm{diag} \big(e^{j\varphi_{1,1}\left[n \right]},
\cdots, e^{j\varphi_{S_{\mathrm{R}_{x}},S_{\mathrm{R}_{y}}}\left[n \right]}\big)\in \mathbb{C}^{S_\mathrm{R}\times S_\mathrm{R}}$ donates the phase shift matrix of the IRS and $\varphi_{S_{\mathrm{R}_{x}},S_{\mathrm{R}_{y}}}\left[n \right]\in\left[0,2\pi \right) $ refers to the phase control introduced
to the $\big(S_{\mathrm{R}_{x}},S_{\mathrm{R}_{y}} \big)$-th reflecting element.
Once the MUAV sends modulated symbol $x^{\mathrm{Tx}}_{k}\left [ n \right ]$  with zero
mean and unit variance at time slot $n$, the received signal at GN $k$ via reflected path can be written as
\begin{align}
	y^{\mathrm{Rx}}_{k}\left [ n \right ]= p_{t}[n]\mathbf{h}_{\mathrm{MG}_k}\left [ n \right ] ^{H}\mathbf{w}_{k}\left [ n \right ] x^{\mathrm{Tx}}_{k}\left [ n \right ]+z_{k}\left [ n \right ],
\end{align}
where $\mathbf{w}_{k}\left [ n \right ] \in \mathbb{C}^{S_\mathrm{M}\times 1}$ is a unit-power beamformer adopted by the MUAV to serve GN $k$ and  $p_{t}[n]$ donates the  information transmit power. The noise at the receivers follows the circular complex Gaussian distribution, i.e., $z_{k}\left [ n \right ] \sim  \mathcal{CN}(0, \sigma^2 )$, where $\sigma^2$ refers to noise power. For simplicity, we assume maximum ratio transmission (MRT) is adopted for power allocation at the transmitter, i.e. $\mathbf{w}_{k}\left [ n \right ]=\frac{1}{\sqrt{S_{\mathrm{M}}}} \mathbf{a}\left(\frac{\pi}{2}-\theta_{\rm{MR}}^{\rm{D}}\left [ n \right ], \xi_{\rm{MR}}^{\rm{D}}\left [ n \right ], S_{\rm{M}_x},S_{\rm{M}_y} \right)$. Then the
optimal phase shift control at the IRS can be directly derived in terms of AoAs/AoDs:
\begin{align}	
	\varphi_{s_{\mathrm{R}_{x}},s_{\mathrm{R}_{y}}}\left[n \right]=\frac{2 \pi \Delta d} {\lambda_{c}}\big[ \left(s_{\mathrm{R}_{x}}\!-1\! \right) \big( \sin\theta_\mathrm{MR}^A\left [n\right ]\cos\xi _\mathrm{MR}^A\left [n\right ]\nonumber\\
	+\sin\theta_{RG_k}^D\left [n\right ]\cos\xi _\mathrm{RG_k}^D \left [n\right ]\big)-\big(s_{\mathrm{R}_{y}}\!-1\! \big)\big( \sin\theta_\mathrm{MR}^A\left [n\right ]\nonumber\\ \sin\xi _\mathrm{MR}^A\left [n\right ]
	+\sin\theta_{RG_k}^D\left [n\right ]\sin\xi _\mathrm{RG_k}^D \left [n\right ]\big) \big], \forall s_{\mathrm{A}_{x}},s_{\mathrm{A}_{y}},n \label{eq:IRS_phase}.
\end{align}
The derivation details are available in Appendix~\ref{appendix},  and the achievable data rate of GN $k$ at time slot $n$ can be calculated as
\begin{align}
	R_{k}\left[n \right]=\log_2\left( 1+\frac{p_{t}[n] \left\|\left(\mathbf{h}_\mathrm{MG_k}\left [ n \right ] \right)^{H}\mathbf{w}_{k}\left [ n \right ]\right\|^2 }{\sigma^2}\right)\label{eq:data_rate}.
\end{align}

\subsection{Energy Transmission Model}
The received power at the AUAV at time slot $n$ can be written as 
\begin{align}
	p_r[n] = p_c \left\|\mathbf{H}_\mathrm{MA}\left [ n \right ]\mathbf{w}_{\rm{M}}\left [ n \right ]\right\|^2,
\end{align}
where the steering vector $\mathbf{w}_{\rm{M}}\left [ n \right ] \in \mathbb{C}^{S_\mathrm{M}\times 1}$ denotes the phase shift vector of UPA carried by the MUAV and constant $p_c$ the charging power transmitted by the MUAV. Similarly, we also adopt MRT for power allocation, i.e. $\mathbf{w}_{\rm{M}}\left [ n \right ]=\frac{1}{\sqrt{S_{\mathrm{M}}}}\mathbf{a}\left(\theta_{\rm{MA}}^{\rm{D}}\left [ n \right ], \xi_{\rm{MR}}^{\rm{D}}\left [ n \right ], S_{\rm{M}_x},S_{\rm{M}_y} \right)$. Then the received power can be simplified as
\begin{align}
	p_r[n] =  S_\mathrm{M}S_\mathrm{A}p_c\alpha_{\rm{MA}}[n].
\end{align}

\subsection{Quad-Rotor UAV Energy Consumption Model}
In this paper, we consider a quad-rotor UAV propulsion consumption model which contains three parts: the blade profile power,  the parasite power and the induced power to overcome the profile drag of blades, fuselage drag, and induced drag, respectively. All of these are related to the velocity of UAV $v$ as mentioned in \cite{8663615}:
\begin{align}
	P_{\mathrm{UAV}}\left(v \right) &=\underbrace{P_{0}\left(1+\frac{3v^2}{\Omega^2R^2} \right) }_{\text{Blade profile}} +\underbrace{P_{i}\left( \sqrt{1+\frac{v^4}{4v_{0}^4}}-\frac{v^2}{2v_{0}^2}\right) ^\frac{1}{2}}_{\text{Induced}}\nonumber\\
	&+\underbrace{\frac{1}{2}d_{0}\rho sA v^{3}}_{\text{Parasite}}. \label{eq:UAVconsumptionmodel}
\end{align}
The physical meanings of the parameters in \eqref{eq:UAVconsumptionmodel} are summarized in Table~\ref{tab1}. Besides, the power for policy computing and IRS controller is negligible compared with the propulsion consumption, which are not considered. Thus, we only take the energy consumption of propulsion, communication and charging into account.

\begin{table}[t]\scriptsize
	\caption{Physical meaning of parameters in Quad-Rotor UAV Energy Consumption Model}
	\setlength{\tabcolsep}{1mm}{
		\begin{tabular}{c|c|c}\rowcolor{gray!40}\hline
			Parameters&Physical meanings&Simulation values\\\hline
			$\Omega$&Blade angel velocity&300 (radians/second)\\
			$R$&Rotor radius&0.4 (meter)\\
			$\rho$&Air density&1.225 (kg/m\textsuperscript{2})\\
			$s$&Rotor solidity&0.05 (m\textsuperscript{3}) \\
			$A$&Rotor disc area&0.503 (m\textsuperscript{2})\\
			$v_{0}$&Induced velocity for rotor in forwarding flight&4.03 (meter/second)\\
			$d_{0}$&Fuselage drag ratio&0.6\\
			$P_{0}$&Blade profile power in hovering status&79.86 (watt) \\
			$P_{i}$&Induced power in hovering status&88.63 (watt)\\\hline
	\end{tabular}}
	\label{tab1}
\end{table}	

\section{Task Description and Problem Formulation}\label{task_description_p_form}
\subsection{Task Description}\label{task_description}
In our proposed scenario, two UAVs leave the initial charging station and provide communication services for GNs in sequence, and do not return until the task termination. Thus, the AUAV with low-capacity battery will power off before the MUAV. We consider two cases here: 1) Both UAVs terminate the task and return to the charging station. 2) The AUAV harvests energy from the MUAV to continue the task. In other words, not only do UAVs have to decide when to  communicate, charge and terminate, i.e., over-option policy, but also how to execute them, i.e., intra-option policy. Given the low time proportion of the latter two processes, we replace their intra-option policies with fixed strategies in the sense of energy minimization, since the optimization makes little sense. Specifically, the two UAVs will approach each other at the max-energy-efficiency velocity, i.e., $v_{mee}=\arg \underset{v}{\max} (\frac{v}{P_{\rm{UAV}}(v)})$,  and hover when the horizontal positions overlap in the process of charging, while for the termination process, two UAVs will head straight to the charging station at $v_{mee}$. Note that the termination process performs multi-time-step execution dependent on $d_{\mathrm{MC}}[n]$ and $d_{\mathrm{AC}}[n]$, while the other two are executed step by step. For the sake of latter presentation, we introduce the concept of option to represent the process to be performed. Denote $\omega^n\in\left\lbrace0, 1, 2 \right\rbrace $ as the current option, referring to the processes of charging, communication and task termination, respectively.  
Then the leftover energy of MUAV and AUAV at time slot $n$ can be formulated as
\begin{align}
	E_\mathrm{M}\left [n\right ]= E_{\mathrm{M}_{max}}\!\!-\delta _{\tau}\sum_{z=0}^{n-1}(P_{\mathrm{UAV}}\left(v_{\mathrm{M}}\left[z \right]\right)+(1-\xi_d)\omega^np_t[z])\nonumber\\-\xi_d E_{\mathrm{Mr}}[n] ,\\
	E_\mathrm{A}\left [n\right ]=\min \big( E_{\mathrm{A}_{max}}, E_{\mathrm{A}_{max}}-\delta _{\tau}\sum_{z=0}^{n-1}(P_{\mathrm{UAV}}\left(v_{\mathrm{A}}\left[z \right]  \right) \nonumber\\ -(1-\xi_d)(1-\omega^n)\alpha_{c}p_r[z])-\xi_dE_{\mathrm{Ar}}[n]\big),
\end{align}
where $v_{\mathrm{M}}[z], v_{\mathrm{A}}[z]$ donate the time-varying velocities of the MUAV and AUAV, respectively, $\alpha_{c}$ donates the energy conversion efficiency and $\xi_d$ is a binary variable masking the termination step, i.e. $\omega^{N_\tau}=2$. Note that the remaining energy of AUAV after charging cannot exceed the battery capacity and $E_{ir}[n]$ donates the energy for return, given by
\begin{align} 
	E_{ir}[n] = \delta_{\tau}P_{\rm{UAV}}(v_{mee})\frac{d_{i\mathrm{C}}[n]}{v_{mee}}, i=\rm{M}, \rm{A}.
\end{align}
Besides, we also denote 
\begin{align}
	E_{il}[n] = E_i[n] - E_{ir}[n], i=\rm{M},\rm{A}, 
	\label{eq:conditional_leftover_energy}
\end{align}
as the conditional leftover energy if return to the charging station at time slot $n$.

\subsection{Problem Formulation}
In this paper, our target is to maximize the accumulative throughput of GNs by jointly optimizing the over-option policy and intra-option policy, including the trajectories of two UAVs and the transmit power at the MUAV. Particularly, both the non-charging and charging scenarios are considered.
\subsubsection{Non-charging Scenario}
That is, $\omega^n=1, \forall n\in\left\lbrace 0,\cdots,N_\tau-1 \right\rbrace$. Besides, the moving range is a rectangular area with the $x$-dimensional limitation $\left[ x_{min}, x_{max} \right] $ as well as the $y$-dimensional limitation  $\left[ y_{min}, y_{max} \right] $. Both UAVs can move freely in the specified area and adjust velocities dynamically. Similarly, the velocity and the transmit power for communication also have the bound $v_{max}, p_{tmax}$. Meanwhile, the UAV crash due to battery exhaustion is not expected at any time, even on the way back to the charging station. So the task ends when the constraint (\ref{constaint1:energy}) is about to be violated.
In practice, signals cannot be reflected when the MUAV and GN $k$ are at different sides of the IRS, so we set the same-side constraint (\ref{constaint1:same_side}) to avoid it. Moreover, we also consider the limited communication coverage capability of the MUAV in the constraint (\ref{constaint1:coverage_range}). For the convenience of illustration, denote $\mathbf{L}=\left\lbrace \textbf{l}_\mathrm{M}\left [ n \right ], \textbf{l}_\mathrm{A}\left [ n \right ] ,\forall n\right\rbrace$ as the trajectory set of two UAVs and $\mathbf{P}_t=\left\lbrace p_{t}[n], \forall n\right\rbrace$ the communication power set allocated to the GNs. Then the problem (P1) can be mathematically formulated as
\begin{subequations}
	\label{P1} 
	\begin{align}
		&\text{(P1)}: \underset{\mathbf{L},\mathbf{P}_t}{\max}  \quad \delta_{\tau}\sum_{n=0}^{N_{\tau}}(1-\xi_d)\omega^nR_{k}\left[n \right], \\
		&\text{s.t.}\quad x_{\rm{M}}[n] \in [x_{min}, x_{max}], x_{\rm{A}}[n] \in [x_{min}, x_{max}],\label{constaint1:xlim}\\ 
		&\quad \quad \;  y_{\rm{M}}[n] \in [y_{min}, y_{max}], y_{\rm{A}}[n] \in [y_{min}, y_{max}],\label{constaint1:ylim}\\  
		&\quad \quad \; v_{\rm{M}}[n] \in [0, v_{max}], v_{\rm{A}}[n] \in [0, v_{max}],\label{constaint1:velocity}\\ 
		&\quad \quad \; p_t[n] \in [0, p_{tmax}],\label{constaint1:transmit_power}\\ 
		&\quad \quad \;\;\textbf{l}_\mathrm{M}\left [ 0 \right ]=\textbf{l}_\mathrm{A}\left [ 0 \right ]=\textbf{l}_\mathrm{M}\left [ N_\tau \right ]=\textbf{l}_\mathrm{A}\left [ N_\tau \right ]=\textbf{l}_\mathrm{C}, \label{constaint1:destination}\\
		&\quad \quad \;E_{il}[n] > 0, i=\rm{M},\rm{A}, \label{constaint1:energy}\\
		&\quad \quad  \;\left( y_{\rm{M}}[n]-y_{\rm{A}}[n] \right)\cdot\left( y_{G_k}[n]-y_{\rm{A}}[n] \right)>0,\label{constaint1:same_side}\\ 
		&\quad \quad \;\left\| \textbf{l}_{\mathrm{M}}[n]-\textbf{l}_{\mathrm{G}_{k}}[n]\right\|\leq l_{min}.\label{constaint1:coverage_range}
	\end{align}
\end{subequations}
\subsubsection{Charging Scenario}
Given the influence of IRS weight on the loading capacity, the battery capacity of AUAV is significantly reduced. A UPA is applied to the top of AUAV for energy harvesting from the MUAV. Besides the optimization variables mentioned in P1, we also focus on the option selection to ensure flight endurance and successful return. Note that the constraint (\ref{constaint2:energy}) is defined in terms of the current leftover energy instead of $E_{il}[n]$, since the charging mechanism is added. Besides, the constraints (\ref{constaint2:same_side}) and (\ref{constaint2:coverage_range}) only restrain the communication slots, i.e. $\omega^n=1$.
Denote $\Omega_\omega=\left\lbrace \omega^n, \forall n \right\rbrace $ as the option set selected by UAVs. Then the problem (P2) can be formulated as
\begin{subequations}
	\label{P2} 
	\begin{align}
		&\text{(P2)}: \underset{\mathbf{L}, \mathbf{P}_t, \Omega_\omega}{\max}  \quad \delta_{\tau}\sum_{n=0}^{N_{\tau}}(1-\xi_d)\omega^nR_{k}\left[n \right], \\
		&\text{s.t.}\quad (\ref{constaint1:xlim}), ({\ref{constaint1:ylim}}), ({\ref{constaint1:velocity}}), ({\ref{constaint1:transmit_power}}),({\ref{constaint1:destination}}), \nonumber\\
		&\quad \quad \; E_{i}[n]>0, i=\rm{M},\rm{A}, \label{constaint2:energy}\\
		&\quad \quad  \;(1-\xi_d)\omega^n\cdot \left( y_{\rm{M}}[n]-y_{\rm{A}}[n] \right)\cdot\left( y_{G_k}[n]-y_{\rm{A}}[n] \right)\geq0,\label{constaint2:same_side}\\ 
		&\quad \quad \;(1-\xi_d)\omega^n\cdot\left\| \textbf{l}_{\mathrm{M}}[n]-\textbf{l}_{\mathrm{G}_{k}}[n]\right\|\leq l_{min}.\label{constaint2:coverage_range}
	\end{align}
\end{subequations}

\section{Multi-Agent DRL for P1}\label{madrl}
The optimization problem (P1) is challenging since it needs a joint design of the UAVs' trajectories and the transmit power, which are highly coupled due to the energy consumption. Thus, in this section, we propose an algorithm named CT-MADDPG to solve this problem.

\subsection{Markov Game for Multi-UAV Cooperation}\label{markovgame}
In a multi-agent environment, the decision making by an agent is influenced by others'. We use the Markov Game as a multi-agent extension of Markov decision process (MDP) to model it. A Markov Game can be represented by a tuple $\left(\mathcal{S},\mathcal{A},\mathcal{R},\mathcal{P},\gamma \right)$, where $\mathcal{S}$ is the  state space, $\mathcal{A}$ is the action space, $\mathcal{R}$ is the reward space, $\mathcal{P}$ is the transition probability, and $\gamma\in[0, 1]$ is the reward discount factor. At time slot $n$, agent $i$ observes the current state $\textbf{s}_{i}^{n}\in\mathcal{S}$ and takes an action $\textbf{a}_{i}^{n}\in\mathcal{A}$ based on its own policy. Then the environment transits into the next state $\textbf{s}_{i}^{n+1}$ with the transition probability $\mathcal{P}\left(\textbf{s}_{i}^{n+1}|\textbf{s}_{i}^{n},\textbf{a}_{1}^{n},\cdots,\textbf{a}_{L}^{n} \right) $ and generates a feedback in term of timely reward $r_{i}^{n}$. Each agent is aimed at maximizing the accumulative discounted rewards, given by
\begin{align}
	R_i^{acc}=\sum_{n=0}^{N_{\tau}}\sum_{m=0}^{N_{\tau}-n}(\gamma)^{m} r_i^{n+m}, 
\end{align}
where $N_{\tau}$ is the time horizon aforementioned.

\begin{figure}[t]
	\centering
	\includegraphics[height=0.9\columnwidth, width=0.85\columnwidth] {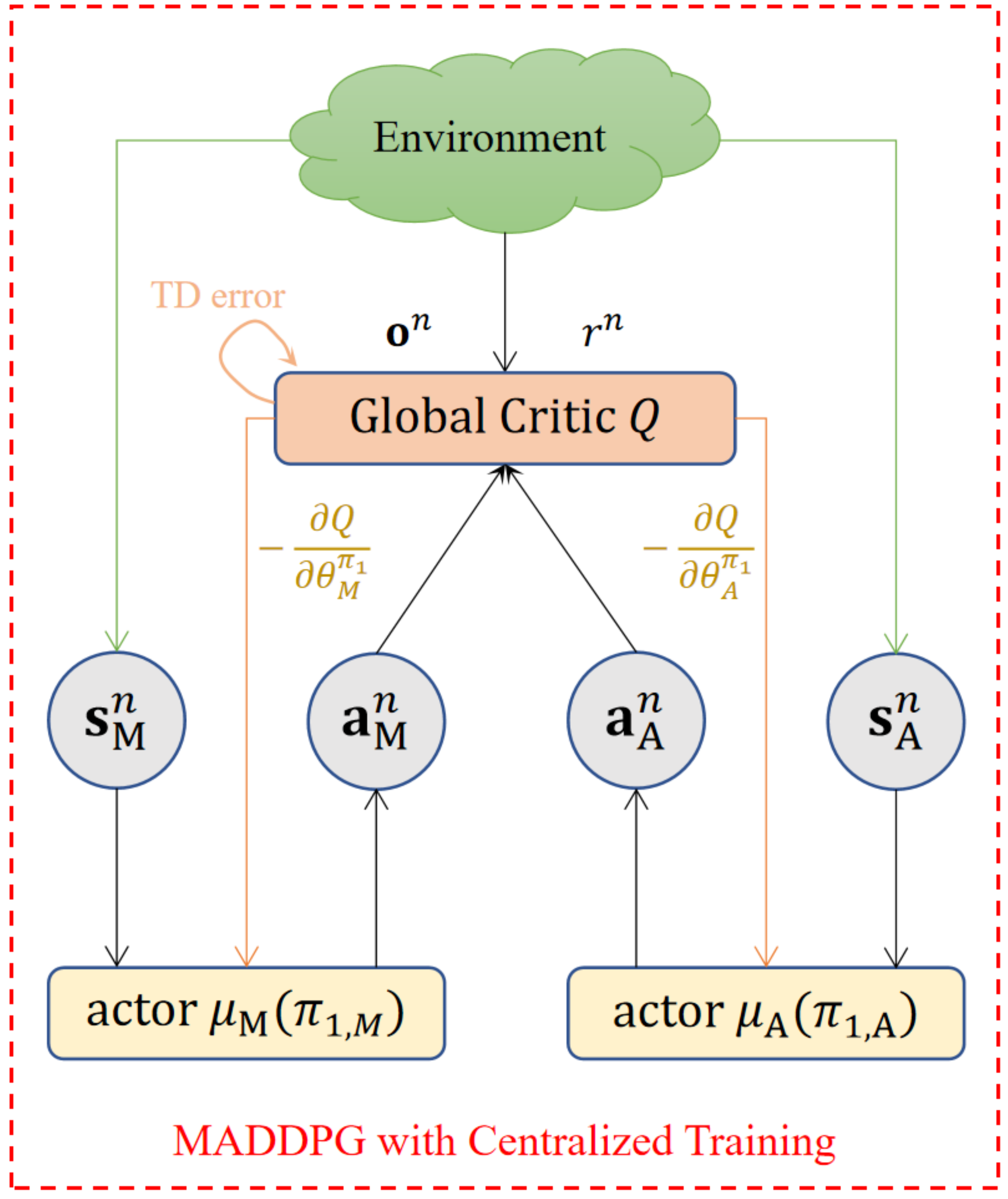}
	\caption{The architecture of CT-MADDPG.}
	\label{fig:MADDPG_architecture}
\end{figure}

\subsection{CT-MADDPG Approach}\label{maddpg}
MADDPG is a multi-agent DRL method applied in multi-agent continuous action space within actor-critic framework. Its main idea is adopting the deep neutral network (DNN), also known as function approximator, to approximate the deterministic policy $\mu: \mathcal{S}\rightarrow\mathcal{A}$ and critic the action executed in a certain state by a fitting function $Q$ for each agent \cite{9043893}. To wherever possible reduce the number of network parameters that need to be optimized, as shown in Fig~\ref{fig:MADDPG_architecture}, we adopt the mode of centralized training, in which all agents share a single reward $r^{n}$ and a global critic network $Q$ is used to evaluate all executed actions in a global observation $\textbf{o}^n$ containing all environment information. We call the MADDPG algorithm with the centralized training method CT-MADDPG. Note that $\textbf{o}^n=\textbf{s}_i^n$ in general except in partially observable environments \cite{9453825}, where agents can only observe local states caused by the limited communication capabilities. Besides, each agent is assigned an actor network $\mu_i$ and executes the action in a distributed way.

Due to the divergence caused by the non-linear function approximator, two techniques are usually used to resolve this issue: the experience replay and the target network. In our scenario, CT-MADDPG stores both agents' experience transitions $\left\lbrace \left\langle \textbf{o}^n, \textbf{s}^{n}_{i},\textbf{a}^{n}_{i},r^{n},\textbf{o}^{n+1},\textbf{s}^{n+1}_{i} \right\rangle,i=\rm{M},\rm{A}\right\rbrace $  into a replay buffer in the exploration stage and randomly samples a mini-batch from it during the training phase. Moreover, the target networks of actor $i$ and global critic,  $\mu'_i$ and $Q'$, which are used to compute the updated target, have the same architectures as the evaluation networks  $\mu_i$ and $Q$. Specifically, the global critic network can be trained by minimizing the temporal difference (TD) error:
\begin{align}
	L\left(\theta^{Q} \right) = \frac{1}{2}\left( y-Q\left(\textbf{o}^{n},\mathbf{a}^{n};\theta^{Q} \right) \right)^2,\label{eq:losscritic}
\end{align}
where
\begin{align}
	y=r^{n}+\gamma Q'\left(\textbf{o}^{n+1},\mathbf{a}^{n+1};\theta^{Q'} \right) * \left( 1-\xi_{d}\right) \label{targetupdate}
\end{align}
is the update target. Note that $\textbf{a}^{n}= \left\lbrace \textbf{a}_{\rm{M}}^{n}, \textbf{a}_{\rm{A}}^{n}\right\rbrace $ donates the actions executed by two agents, $\textbf{o}^{n+1}$ donates the next global observation and $\textbf{a}^{n+1}=\big\lbrace \mu'_i\big(\mathbf{s}^{n+1}_{i};\theta^{\mu'}_{i}\big),i=\rm{M},\rm{A}\big\rbrace $ is given by the target actor networks  of both agents. Besides, $\theta^{Q}$ and $\theta^{Q'}$ refer to the parameters of evaluation and target critic networks as well as the evaluation and target actor networks are weighted by $\theta^{\mu}_{i}$ and $\theta^{\mu'}_{i}$, respectively. Moreover, the actor network of agent $i$ is trained by minimizing the Q value:
\begin{align}
	L\left(\theta^{\mu}_{i} \right) = -Q\left(\textbf{o}^{n},\mu_i\left(\textbf{s}^{n}_{i};\theta^{\mu}_{i} \right);\theta^{Q} \right).\label{eq:lossactor}
\end{align}
In addition, the target networks are updated by slowly tracking the learned networks:
\begin{align}
	\theta^{\mu'}_{i} &= \left(1-\tau \right) \theta^{\mu'}_{i} + \tau\theta^{\mu}_{i},\label{eq:targetactorupdate}\\
	\theta^{Q'} &= \left(1-\tau \right) \theta^{Q'} + \tau\theta^{Q},\label{eq:targetcriticupdate}
\end{align}
where $\tau \ll 1$ donates the soft replace rate. The other two tricks of CT-MADDPG are Gaussian action noise $\mathcal{N}_\mathcal{A}$ addition to ensure the exploratory of agents and noise attenuation in the late training period to achieve better convergence.

\subsection{Space Design}
In this subsection, we give a rational design including the spaces of state, action and reward for (P1). 

\textbf{State:} The states observed by the MUAV and the AUAV at time slot $n$ are the same one-dimensional vectors with nine components:
\begin{itemize}
	\item[$\bullet$] $\textbf{l}_\mathrm{M}\left [ n \right ], \textbf{l}_\mathrm{A}\left [ n \right ]$: the horizontal positions of two UAVs.
	\item[$\bullet$] $H_{\mathrm{M}},H_{\mathrm{A}}$: the operation altitudes of two UAVs.
	\item[$\bullet$] $\textbf{l}_{\mathrm{G}_{k}}$: the horizontal position of GN $k$.
	\item[$\bullet$]$E_\mathrm{M}\left [n\right ], E_\mathrm{A}\left [n\right ]$: the leftover energy of two UAVs.
	\item[$\bullet$]$E_{\mathrm{M}l}\left [n\right ], E_{\mathrm{{A}l}}\left [n\right ]$: the conditional leftover energy defined in (\ref{eq:conditional_leftover_energy}).
\end{itemize}
Note that the location information can be observed through additional GPS devices, and the electricity information can be shared with simple signal communications, which are readily available in practice. Hence, the full state can be formulated as
\begin{align}
	\textbf{s}_i^n=  \big\lbrace \textbf{l}_\mathrm{M}\left [ n \right ], H_{\mathrm{M}}, \textbf{l}_\mathrm{A}\left [ n \right ], H_{\mathrm{A}}, \textbf{l}_{\mathrm{G}_{k}},E_\mathrm{M}\left [n\right ],E_\mathrm{A}\left [n\right ],\nonumber\\ E_{\mathrm{M}l}[n], E_{\mathrm{A}l}[n]\big\rbrace ,i=\rm{M},\rm{A}, \label{observation_information}
\end{align}
and has $12$ dimensions. Moreover, the global observation $\textbf{o}^n = \textbf{s}_{i}^n$ specifically in our scenario. 

\textbf{Action:} The actions for the MUAV and the AUAV at time slot $n$ donated by $\textbf{a}_{\rm{M}}^n$ and $\textbf{a}_{\rm{A}}^n$ are the outputs of their actor networks that contain the scalar elements shown in the following: 
\begin{itemize}
	\item[$\bullet$] $v_{\mathrm{M}}\left[n \right],v_{\mathrm{A}}\left[n \right]$: The MUAV's and the AUAV's velocity.
	\item[$\bullet$] $a_{\mathrm{M}}\left[n \right],a_{\mathrm{A}}\left[n \right]\in \left[-\pi,\pi \right] $: The azimuthal angles.
	\item[$\bullet$] $p_t\left[n \right]$: The MUAV's transmit power.
\end{itemize}

According to the above definitions, the action sets of the MUAV and the AUAV can be written as 
\begin{align}
	\textbf{a}_{\rm{M}}^n&=\left[v_{\mathrm{M}}\left[n \right], a_{\mathrm{M}}\left[n \right], p_t\left[n \right]\right],\label{eq:action_MUAV}\\
	\textbf{a}_{\rm{A}}^n&=\left[v_{\mathrm{A}}\left[n \right], a_{\mathrm{A}}\left[n \right]\right]\label{eq:action_AUAV}.
\end{align}

\textbf{Reward:} In the centralized training mode, there is a global reward for agents sharing,  which depends on the current global state and all executed actions. In our scenario, the reward design includes two parts:
\begin{itemize}
	\item[$\bullet$]Single-step throughput reward: First, we need to motivate both UAVs by including a single-step throughput reward in a shared manner in order to solve (P1). The reward is designed as 
	\begin{align}
		r_{th}^n=\xi_{th}[n]\omega^n\left( R_{k}\left[n \right]\right) ^{\kappa_{th}}, 
	\end{align}
	where $\kappa_{th}$ is a factor to adjust the distribution of the reward. $\xi_{th}[n ]=0$ if violating the constraint  (\ref{constaint1:same_side}) or (\ref{constaint1:coverage_range}), otherwise,  $\xi_{th}[n]=1$.
	\item[$\bullet$]Cross-border penalty: To prevent the UAVs from leaving the operation area, we set the cross-border penalty as
	\begin{align}
		r_{cb}^n=\kappa_{cb}\left( \xi_{Mcb}\left[n \right]+ \xi_{Acb}\left[n \right]\right),
	\end{align}
	where $\xi_{Mcb}[n]$ and $\xi_{Acb}[n]$ are binary indicators implying whether the MUAV and the AUAV are outside the square area or not, respectively. Besides, $\kappa_{cb}$ is a negative constant.
\end{itemize}

Thus, the overall  global reward at time slot $n$ is formulated as
\begin{align}                                                           
	r^n=r_{cb}^n+r_{th}^n.
\end{align}

\section{option-based Multi-Agent HDRL for P2}\label{mahdrl}
The CT-MADDPG method aforementioned only performs as a surrogate of the intra-option policy, while in the charging scenario, an extra higher-level surrogate of the over-option policy is needed. To deal with this issue, we combine the CT-MADDPG with the options framework \cite{OC} to enable the multi-agent hierarchical control.

\subsection{Semi-Markov Game for Multi-UAV Options Learning}
\textbf{Options:} The options framework allows a reinforcement learning agent to represent, learn and plan with temporally extended action. The temporally extended action, also known as Markovian option $\omega$, is usually represented by a triplet $\left\langle I_{\omega}, \pi_{\omega}, \beta_{\omega}\right\rangle $, in which $I_{\omega} \subseteq \mathcal{S}$ is an initial set, $\pi_{\omega}$ is an intra-option policy, and $\beta_{\omega}: \mathcal{S} \rightarrow [0,1]$ is a termination function. If $\textbf{s}^n \in I_{\omega} $, then the option $\omega^n$ is available at state $\textbf{s}^n$. Once an option is selected, a series of corresponding intra-option actions will be executed based on $\pi_{\omega}$ until it terminates with $\beta_{\omega}: \mathcal{S} \rightarrow \mathcal{A}$, then the over-option policy  $\pi_{\Omega}: \mathcal{S} \rightarrow \Omega$ will make a new choice, at which point this procedure is repeated.

\textbf{SMDP:} Distinguished from the mode of single-step action execution in the traditional MDP, the states may change several times between two adjacent decisions in SMDP, while the intermediate states may be irrelevant to the agent. Moreover, the intra-option policy $\pi_{\omega}$ can be even replaced  with the artificial policy served as a sequence of pre-specified actions. Hence, the SMDP can be regarded as a special application of MDP in temporal abstract scenario, where the time step of each policy execution is a random variable.

As we introduce the options framework to the system, the option set $\Omega_{\omega}$ should also be added to the Markov game tuple. Hence, we can use semi-Markov game, i.e. $\left\langle \mathcal{S},\Omega_{\omega}, \mathcal{A},\mathcal{R},\mathcal{P},\gamma\right\rangle  $, to model (P2). Note that the options serve as global decisions for both agents specifically in our scenario, since whether the communication or the charging process can not be finished by a single UAV, and also there is no guarantee that both UAVs will make the same choice simultaneously.

\subsection{MADDPOC}
\begin{figure}[t!]
	\centering
	\includegraphics[height=0.9\columnwidth, width=0.85\columnwidth] {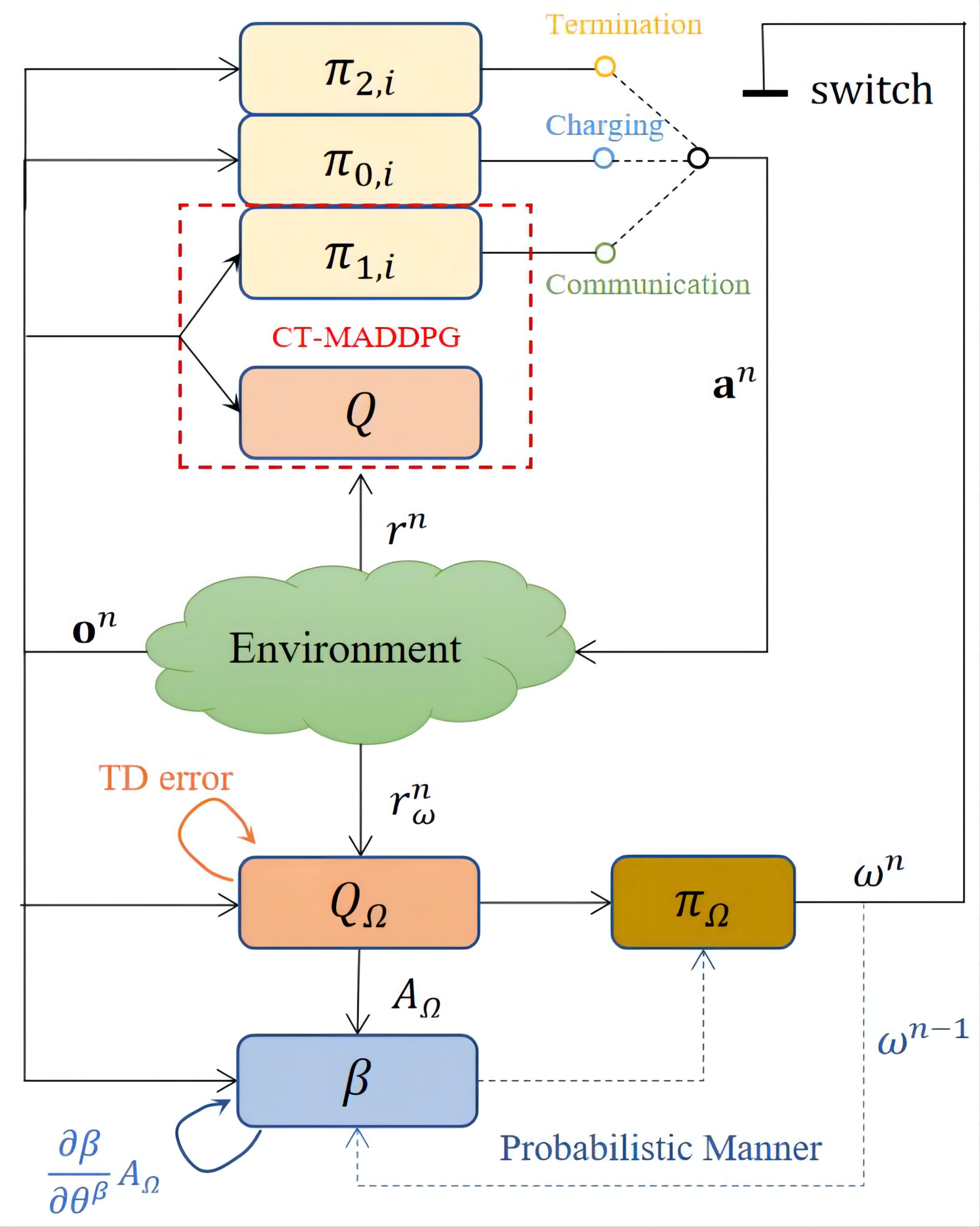}
	\caption{The architecture of MADDPOC.}
	\label{fig:MADDPOC_architecture}
\end{figure}

We focus on the global options learning for both agents in this subsection. To achieve this goal, the CT-MADDPG algorithm is combined with the options framework by adding two sets of independent differentiable parameterized
function approximators. As shown in Fig~\ref{fig:MADDPOC_architecture}, $\pi_{\omega, i}$ denotes the agent $i$'s intra-option policy of option $\omega$ parameterized by $\theta_i^{\pi_\omega}$, $\beta$ the termination function of options parameterized by $\theta^{\beta}$, and $Q_{\Omega}$ the global state-option value function parameterized by $\theta^{Q_{\Omega}}$. Besides, donate $\beta'$, $Q'_{\Omega}$ weighted by $\theta^{\beta'}$, $\theta^{Q'_{\Omega}}$ for the target networks of $\beta$ and $Q_{\Omega}$, respectively.

Just like the critic network evaluates actions, the function $Q_{\Omega}$ is used to evaluate how good or bad different choices are under a particular state. Thus, its outputs can be written as $Q_{\Omega}\left(\textbf{o}^n; \theta^{Q_{\Omega}}\right) \in \mathbb{R}^{1 \times n_{\omega}}$, where $n_{\omega}$ is the total number of available options. And $\beta\left(\textbf{o}^n; \theta^{\beta} \right) \in \mathbb{R}^{1 \times n_{\omega}}$, the outputs of $\beta$ network, donates the termination probability of each option. Similarly, let $Q'_{\Omega}\small(\textbf{o}^{n+1}; \theta^{Q'_{\Omega}}\small)$ and $\beta'\small(\textbf{o}^{n+1}; \theta^{\beta'} \small)$  denote the outputs of their target networks, respectively. For the sake of distinction, we define a function as
\begin{align}
	\mathcal{F} \small(\textbf{o},\omega \small)\triangleq   \sum \mathcal{F}\small(\textbf{o};\theta \small)  \odot \textbf{e}_{\omega}, 
\end{align}
where $\mathcal{F} \in \left\lbrace Q_{\Omega}, \beta, Q'_{\Omega}, \beta' \right\rbrace $ donates some kind of network weighted by $\theta\in \small\lbrace \theta^{Q_{\Omega}}, \theta^\beta, \theta^{Q'_{\Omega}}, \theta^{\beta'}\small\rbrace$, which takes the observation infromation $o$ as input. Note that $\mathcal{F}\small(\textbf{o};\theta \small)$ donates the output vector and $\mathcal{F} \small(\textbf{o},\omega \small)$ the value corresponding to the option $\omega$, which can be formulated as a product sum form of the option's one-hot code $\textbf{e}_{\omega}$ and the network output vector.

As mentioned above that the last option $\omega^{n-1}$ will continue to perform util it terminates, and then a new choice $\omega^{n}$ will be made based on the over-option policy $\pi_{\Omega}$. Two methods are available here:
\begin{align}
	\omega^{n} &= \arg\underset{\omega}{\mathrm{max}}\left(Q_{\Omega}\left(\textbf{o}^n; \theta^{Q_{\Omega}}\right) \right), \label{greedy_pi_omege}\\
	\omega^{n} &=\left\{\begin{matrix}\omega^{n-1}, \text{with probability } 1-\beta\small(\textbf{o}^{n},\omega^{n-1} \small)\\ \!\!\!\!\!\!\!\!\!\!\!\arg\underset{\omega}{\mathrm{max}}\left(Q_{\Omega}\left(\textbf{o}^n; \theta^{Q_{\Omega}}\right) \right), \text{otherwise}
	\end{matrix}\right.\label{probabilistic_pi_omege}
\end{align}
for greedy policy and probabilistic policy, respectively.
Besides, the related update target of $Q_{\Omega}$ takes a mixed form of probability:
\begin{align}
	y_{\Omega}=r^n_{\omega} &+  \gamma \big(  \left(1-\beta'\small(\textbf{o}^{n+1},\omega^n \small)\right)Q'_{\Omega}\small(\textbf{o}^{n+1},\omega^n \small) 
	\nonumber\\ &+  \beta'\small(\textbf{o}^{n+1},\omega^n \small) Q'_{\Omega new}\big)* \left( 1-\xi_{d}\right), 
\end{align} 
where $\beta'\small(\textbf{o}^{n+1},\omega^n \small), Q'_{\Omega}\small(\textbf{o}^{n+1},\omega^n \small)$ donate the termination probability and expected advantage value of option $\omega^n$ related to the next state, respectively. Note that $r^n_{\omega}$ refers to the timely reward feedback given by the environment after the option execution. Besides, the expected advantage of new option  is usually in the form of maximum advantage estimates, i.e. $Q'_{\Omega new}=\max\left( Q'_{\Omega}\small(\textbf{o}^{n+1};\theta^{Q'_{\Omega}} \small) \right) $. To avoid the drawbacks of overestimation, we use the double-Q technique \cite{Double-DQN} instead:
\begin{align}
	Q'_{\Omega new}=Q'_{\Omega}\left(\textbf{o}^{n+1}, \arg\underset{\omega}{\mathrm{max}}\left( Q_{\Omega}\small(\textbf{o}^{n};\theta^{Q_{\Omega}} \small) \right)\right).
\end{align}
Moreover, the network $Q_{\Omega}$ can be trained in the same way as the global critic $Q$ by minimizing the TD error:
\begin{align}
	L\left(\theta^{Q_{\Omega}} \right) = \frac{1}{2} \left( y_\Omega-Q_{\Omega}\left(\textbf{o}^{n}, \omega^n \right) \right)^2.\label{eq:lossQomega}
\end{align}
In addition, the $\beta$ net is updated by minimizing the loss masked by $\xi_{d}$:
\begin{align}
	L\left(\theta^{\beta} \right) =\beta\small(\textbf{o}^{n},\omega^{n-1} \small)A_{\Omega}* \left( 1-\xi_{d}\right),\label{eq:lossBeta}
\end{align}
where $A_{\Omega}=\left(Q_{\Omega}\left(\textbf{o}^{n}, \omega^n \right)-V_{\Omega} \right)$ is a constant, donating the termination advantage value of option $\omega^n$. When the option choice is suboptimal with respect to the expected value $V_{\Omega}$ over all options, the value $A_{\Omega}$ is negative and it drives the gradient corrections up, which increases the odds of terminating. Note that $V_\Omega$ takes the form of max advantage under the greedy policy, i.e. $V_\Omega=\max\left(Q_{\Omega}\small(\textbf{o}^{n};\theta^{Q_{\Omega}} \small) \right) $, or is given in a probabilistic manner:
\begin{align}
	V_\Omega&=\left(1-\beta\small(\textbf{o}^{n},\omega^{n-1} \small)\right)Q_{\Omega}\small(\textbf{o}^{n},\omega^n \small) 
	\nonumber  \nonumber\\ &+\beta\small(\textbf{o}^{n},\omega^{n-1} \small) \max\left( Q_{\Omega}\left(\textbf{o}^{n};\theta^{Q_{\Omega}}  \right) \right) 
\end{align}
Also, the target networks are updated by applying soft replace, which is the same as (\ref{eq:targetactorupdate}) and (\ref{eq:targetcriticupdate}):
\begin{align}
	\theta^{Q'_{\Omega}} &= \left(1-\tau \right) \theta^{Q'_{\Omega}} + \tau\theta^{Q_{\Omega}},\label{eq:targetQomegaupdate}\\
	\theta^{\beta'} &= \left(1-\tau \right) \theta^{\beta'} + \tau\theta^{\beta}.\label{eq:targetBetaupdate}
\end{align}
Different from the policy gradient (PG) method used in \cite{OC}, we no longer design additional networks for intra-option policy learning which can be directly replaced by the CT-MADDPG base, i.e. $\pi_{\omega=1, i} = \mu_i$. In other words, the networks $Q_{\Omega}, \beta$ and $Q, \pi_{1, i}$ maintain two independent experience buffers for the purposes of options learning and intra-option policies optimization, respectively. Besides, the complete training process of MADDPOC is shown in  Algorithm~\ref{AlgorithmMADDPOC}.

\begin{algorithm}[t!]
	\caption{Training process for MADDPOC} \label{AlgorithmMADDPOC}
	\begin{algorithmic}[1]
		\STATE Initialize all the networks, including $Q_{\Omega}, \beta, Q, \pi_{1, i}$ weighted by $\theta^{Q_{\Omega}}, \theta^{\beta}, \theta^{Q}, \theta^{\pi_{1}}_i$, as well as their target nets  $Q'_{\Omega}, \beta', Q', \pi'_{1, i}$. Set target weights: $\theta^{Q'_{\Omega}}=\theta^{Q_{\Omega}},\theta^{\beta'}=\theta^{\beta},\theta^{Q'}=\theta^{Q}, \theta^{\pi'_{1}}_i=\theta^{\pi_{1}}_i$. Note that $i=M, A$ denotes the subscript of the MUAV and AUAV, respectively.
		\STATE Initialize the artificial intra-option policies $\pi_{0, i}, \pi_{2, i}$ weighted by untrainable parameters $\theta^{\pi_{0}}_i$, $\theta^{\pi_{2}}_i$.
		\STATE Initialize the queues $D_{\Omega}, D_{\mathcal{A}}$ with the same capability $C$ as experience replay buffers.
		\FOR{episode = 1 to $\mathcal{M}$}
		\STATE Initialize the primal option $w^{-1}=1$.
		\WHILE{not $\xi_{d}$}
		\STATE Get global observation $\textbf{o}^{n}$ from the environment. 
		\IF {$D_{\Omega}$ is not full}
		\STATE Make choices and execute actions randomly.
		\ELSE
		\STATE Select current option according to (\ref{probabilistic_pi_omege}).
		\STATE Get actions based on the intra-option policy:
		$\textbf{a}^n = \left\lbrace \pi_{\omega^n, i}\left(\textbf{o}^n;\theta^{\pi_{\omega^n}}_i\right)+ (1-\xi_d)\omega^n\mathcal{N_{\mathcal{A}}} , i=M, A\right \rbrace $, where $\mathcal{N_{\mathcal{A}}} \sim  \mathcal{N}(0, \sigma_{\mathcal{A}}^2 ) $ donates the action noise and $\sigma_{\mathcal{A}}^2$ donates the noise power.
		\ENDIF
		\IF{$\textbf{a}_i^n \in \textbf{a}^n$ violates the constraint (\ref{constaint1:transmit_power}) or (\ref{constaint1:velocity})}
		\STATE Clip $\textbf{a}_i^n$ by the action bounds.
		\ENDIF
		\STATE Execute the clipped actions and get feedbacks in terms of the next state $\textbf{o}^{n+1}$ and rewards $r^n, r_\omega^n$.
		\IF{$\textbf{o}^{n+1}$ violates the constraint (\ref{constaint1:xlim}) or (\ref{constaint1:ylim})}
		\STATE Clip $\textbf{o}^{n+1}$ by the area bounds.
		\ENDIF
		\STATE Update the state $\textbf{o}^n = \textbf{o}^{n+1}$.
		\STATE Store $\left\langle \textbf{o}^n,r^n_\omega,\textbf{o}^{n+1},\omega^n,\omega^{n-1}, \xi_{d}\right\rangle$  into $D_\Omega$.
		\IF{$\omega^n=1$}
		\STATE Store $\left\langle \textbf{o}^n,\textbf{a}^n,r^n,\textbf{o}^{n+1}, \xi_{d}\right\rangle$  into $D_{\mathcal{A}}$.
		\ENDIF
		\IF{$D_{\Omega}$ is full}
		\STATE Sample a random minibatch of $\mathcal{B}$ transitions from $D_{\Omega}, D_{\mathcal{A}}$ and train the networks  $Q_{\Omega}, \beta, Q, \pi_{1, i}$ by minimizing the losses \eqref{eq:lossQomega}, \eqref{eq:lossBeta}, \eqref{eq:losscritic} and \eqref{eq:lossactor}, respectively.
		\STATE Update the target networks using the soft replace methods \eqref{eq:targetQomegaupdate},  \eqref{eq:targetBetaupdate}, \eqref{eq:targetactorupdate} and  \eqref{eq:targetcriticupdate}.
		\ENDIF
		\ENDWHILE
		\IF{episode$>\mathcal{M}_{d}$}
		\STATE Decay the action noise in an exponential manner:
		\begin{align}                                              	
			\sigma_{\mathcal{A}}=\sigma_{\mathcal{A}_{0}}\gamma_{\mathcal{A}}^{\frac{\text{episode}-\mathcal{M}_{d}}{\mathcal{M}_{f}}} ,
		\end{align}
		where $\sigma_{\mathcal{A}_{0}}$ donates the initial standard deviation of $\mathcal{N_{\mathcal{A}}}$, $\gamma_{\mathcal{A}}$ denotes the decay rate and $\mathcal{M}_{f}$ refers to the decay frequency.
		\ENDIF
		\ENDFOR
	\end{algorithmic}
\end{algorithm}

\subsection{Space Design}\label{mdddpocspace}
Detailed design on the spaces of state, option, action and reward  for MADDPOC is elaborated in this subsection.

\textbf{State:} All networks take the observation information $\textbf{o}^n$ aforementioned in (\ref{observation_information}) as inputs.

\textbf{Option:} Three options are available in our scenario as mentioned in \ref{task_description}:  $\omega^n=0$ for charging, $\omega^n=1$ for cooperative communication and $\omega^n=2$ for task termination at time slot $n$. Given the serious problem of training sample truncation brought by the termination option, we make it unselectable in the early task. To be specific, we set an energy threshold $E_{th}$ and reformulate the greedy method (\ref{greedy_pi_omege}):
\begin{align}
	\omega^{n} = \arg\underset{\omega}{\mathrm{max}}\left(Q_{\Omega}\left(\textbf{o}^n; \theta^{Q_{\Omega}}\right)-\textbf{q}_{mask} \right), \label{greedy_pi_omege_with_om}
\end{align}
where
\begin{align} 
	\textbf{q}_{mask}=\left\{\begin{matrix}\left[0,0,q_m\right] , &E_{\rm{M}}>E_{th},\\ 
		\left[ 0,0,0\right], &\text{otherwise}.
	\end{matrix}\right.
\end{align}
Note that $q_m$ is a big enough positive number,
and the probabilistic method (\ref{probabilistic_pi_omege}) can be rewritten in the same way. 

\textbf{Action:} We have described the intra-option policies $\pi_{0, i}$ and $\pi_{2, i}$ as charging and termination processes, respectively, in the subsection~\ref{task_description}. In other words, when the agents choose either of the policies, their actions or action sequences to be executed are already artificially determined and no longer the outputs of their actor networks. Besides, CT-MADDPG can serve as a surrogate of $\pi_{1, i}$. Thus, the action space of $\pi_{1, i}$ follows (\ref{eq:action_MUAV}) and (\ref{eq:action_AUAV}).

\textbf{Reward:} Two extra sparse rewards are added which helps to  make options learning accurate and rapidly convergent:
\begin{itemize}
	\item[$\bullet$]Deliberate cost: Charging can ensure the AUAV's endurance, but too much will impact the throughput performance. Also, an early return is a waste of energy. Therefore, we expect reasonable option switching  only when necessary. Inspired by the deliberate cost proposed in \cite{AOC}, a penalty to control the switching frequency is set as
	\begin{align}
		r_{dc}^n=(\xi_d+(1-\xi_d)\omega^{n-1}\left( 1-\omega^n\right)) \kappa_{dc}, 
	\end{align}
	where $\kappa_{dc}$ is a negative constant. Note that only the switch from communication to charging or termination is penalized since the former is what we encourage.
	\item[$\bullet$]Power-off penalty: To avoid the air crash midway due to battery exhaustion, we set a great penalty as
	\begin{align}
		r_{po}^n= \xi_{po}[n]\kappa_{po}, 
	\end{align}
	where $\xi_{po}[n]$ donates a binary indicator representing whether the  constraint (\ref{constaint2:energy}) is violated and $\kappa_{po}$ is a negative constant.
\end{itemize}

\begin{figure*}[t!]
	\centering
	\includegraphics[height=0.9\columnwidth, width=2\columnwidth] {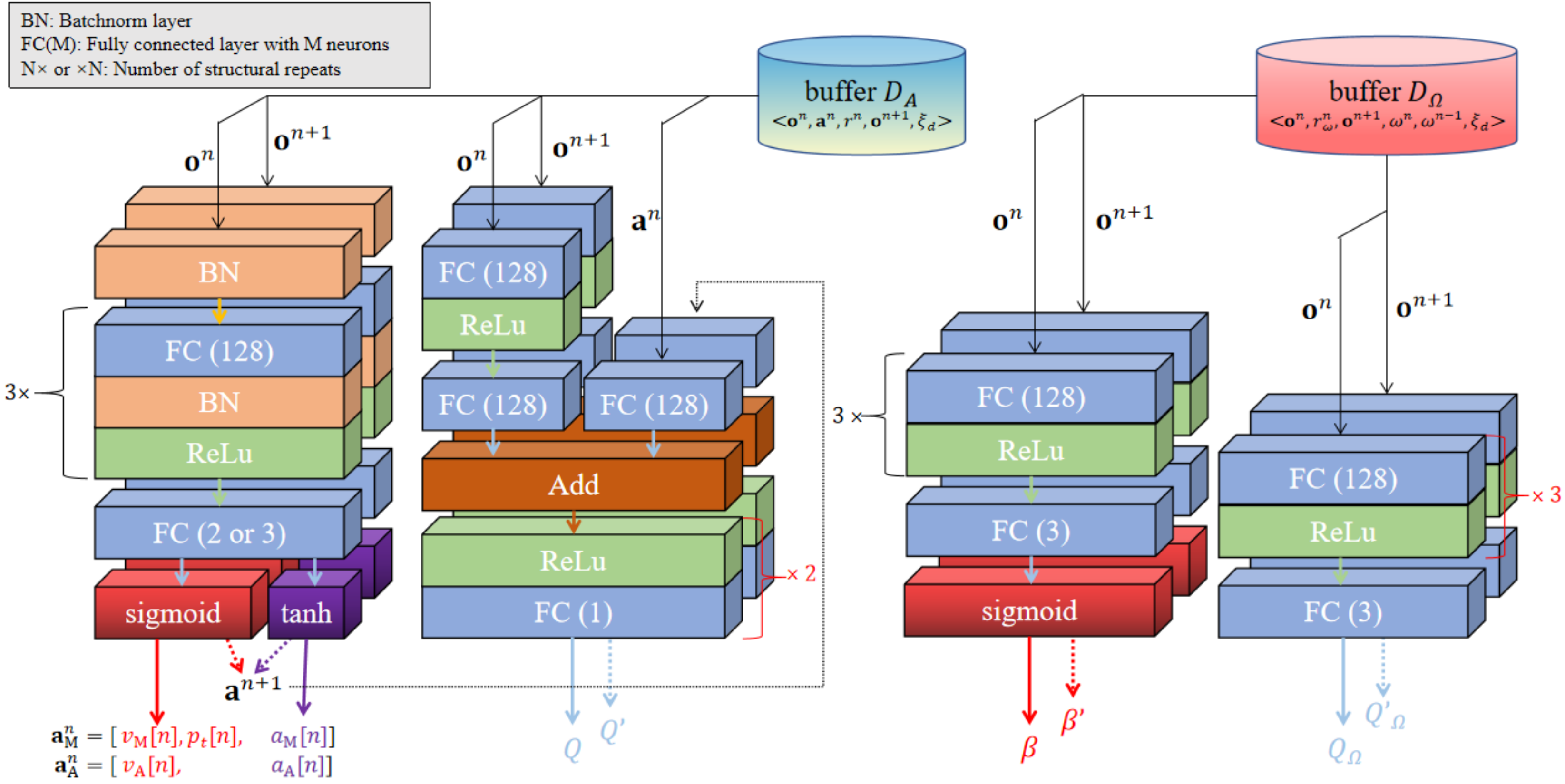}
	\caption{Network structures of $\pi_{1, i}, Q, \beta, Q_{\Omega}$ and their target nets (from left to right).}
	\label{fig:network_settings}
\end{figure*}

In summary, the total reward to compute the update target of $Q_{\Omega}$ at time slot $n$ is formulated as
\begin{align}                                                       
	r^n_\omega=r^n + r_{dc}^n + r_{po}^n.
\end{align}

\section{Numerical Results and Performance Evaluation}\label{result}
In this section, we present the simulation results as well as
the performance analysis of the proposed algorithm.
\subsection{Environment Settings}
Related environment parameters are summarized in the front part of Table~\ref{tab3:settings}. In order to minimize the length of time interval for the termination judgment, we set the threshold  $E_{th}=2p_c\sigma_\tau$, i.e., at least one more chance to charge, since the selection of termination before is a waste of energy and also makes the training samples truncated. Besides, the unspecified parameters, i.e. the number of GNs $K$, and the fully charged energy $E_{\rm{M}_{\rm{max}}}, E_{\rm{A}_{\rm{max}}}$, can be set manually.

\subsection{Training Settings}
We train MADDPOC using the Adam optimizer with
learning rate 0.001. Specific structures of all networks, including $\pi_{1, i}, Q, \beta, Q_{\Omega}$ as well as their corresponding target nets, are shown in Fig~\ref{fig:network_settings}.
Given the inconsistent length of action normalization intervals, we make the initial derivation of action noise $\sigma_{\mathcal{A}_0}$ proportional to the normalized action range, e.g., $\sigma_{N_{v}} = 0.15, \sigma_{N_{a}} = 0.3, \sigma_{N_{p}} = 0.15$ for velocity,
azimuthal angle and communication power, respectively. The rest parameters are shown in the lower half of Table~\ref{tab3:settings}.

\begin{table}\scriptsize
	\caption{Environment and Training Settings}
	\label{tab3:settings}
	\setlength{\tabcolsep}{1mm}{
		\begin{tabular}{c|c|c}\rowcolor{gray!40}\hline
			Parameters&Description&Value\\\hline
			% $f_c$&Carrier frequency&1GHz \\\hline
			$\eta_a, \eta_b, \eta_{\rm{LoS}}, \eta_{\rm{NLoS}}$&Dense Urban Environment&12.08, 0.11, 1.6, 23 \\\hline
			$\sigma^2$& Background noise
			power&$10^{-13}$W/Hz \\\hline
			$(x_{min},x_{max})$&Abscissa interval of the square area&(-100m, 100m) \\\hline
			$(y_{min},y_{max})$&Ordinate interval of the square area&(-100m, 100m) \\\hline
			$H_{\rm{M}}, H_{\rm{A}}$&Flying heights of the MUAV and AUAV&100m, 98m\\\hline
			$\textbf{l}_{\rm{C}}$&Location of the charging station&[-75m, 0m]\\\hline
			$\delta_\tau$&Slot duration&0.5s\\\hline
			$N_u$&Link switching frequency&30\\\hline
			$S_{\rm{M}}, S_{\rm{A}}$&Number of antennas on the UPA&128, 128\\\hline
			$S_{\rm{R}}$&Number of IRS units&128\\\hline
			$p_{tmax}$&Maximum communication power&10W \\\hline
			$p_{c}$&Charge power transmitted by the MUAV&5000W\\\hline
			$v_{max}$&Maximum velocity of UAVs&20m/s\\\hline
			$v_{mee}$&Velocity with max efficiency&18.3m/s\\\hline
			$l_{min}$&Communication coverage distance&25m\\\hline
			$\alpha_{c}$&Energy conversion efficiency&0.9\\\hline	
			$E_{th}$&Leftover energy threshold&5000Joules\\\hline	\hline
			$\gamma$&Reward discount factor&0.95\\\hline
			$\tau$&Soft replace rate&0.001\\\hline
			$\mathcal{B}$&Minibatch size&128\\\hline
			$\mathcal{C}$&Capacity of experience replay buffers&30000\\\hline
			$\mathcal{M}$&Total training epoch&1500\\\hline
			$\sigma_{\mathcal{A}_{0}}$&Initial action noise standard deviation &0.15 or 0.3\\\hline
			$\gamma_{\mathcal{A}}$&Decay rate of the action noise&0.99\\\hline
			$\mathcal{M}_f$&Decay frequency of the action noise&2\\\hline
			$\mathcal{M}_d$&Starting epoch of the noise decay&1000\\\hline
			$\kappa_{th}, \kappa_{cb}, \kappa_{dc}, \kappa_{po}$&Reward settings&1.2, -5, -5, -1000\\\hline
	\end{tabular}}
\end{table}	

\subsection{Performance and Analysis}
We compare MADDPOC with other three baseline methods,
MADDPOC without option masking (MADDPOC-NM), deep deterministic policy option critic (DDPOC) and option critic {OC}, while the CT-MADDPG algorithm proposed for (P1) serves as a contrastive benchmark in the non-charging scenario.
\begin{itemize}
	\item[$\bullet$] MADDPOC-NM: No option masking is applied and all options are assumed to be selectable anytime in the framework of MADDPOC.
	\item[$\bullet$] DDPOC: A single-agent version of MADDPOC, where $\pi_{1, i}$ is replaced by a single actor network $\pi_{1}$ that outputs the actions of both UAVs.
	\item[$\bullet$] OC: The classical hierarchical architecture proposed in \cite{OC}, where $\pi_{1}$ is trained by applying the policy gradient weighted by the advantage value $A_\Omega$.
	
\end{itemize}
The latter two methods apply option masking by default. Unless specified otherwise, our default environment configuration is set to: $E_{\rm{M}_{\rm{max}}}=35000$ Joules, $E_{\rm{A}_{\rm{max}}}=15000$ Joules, and $K=10$.

\begin{figure}[t!]
	\centering
	\includegraphics[height=.8\columnwidth, width=1\columnwidth] {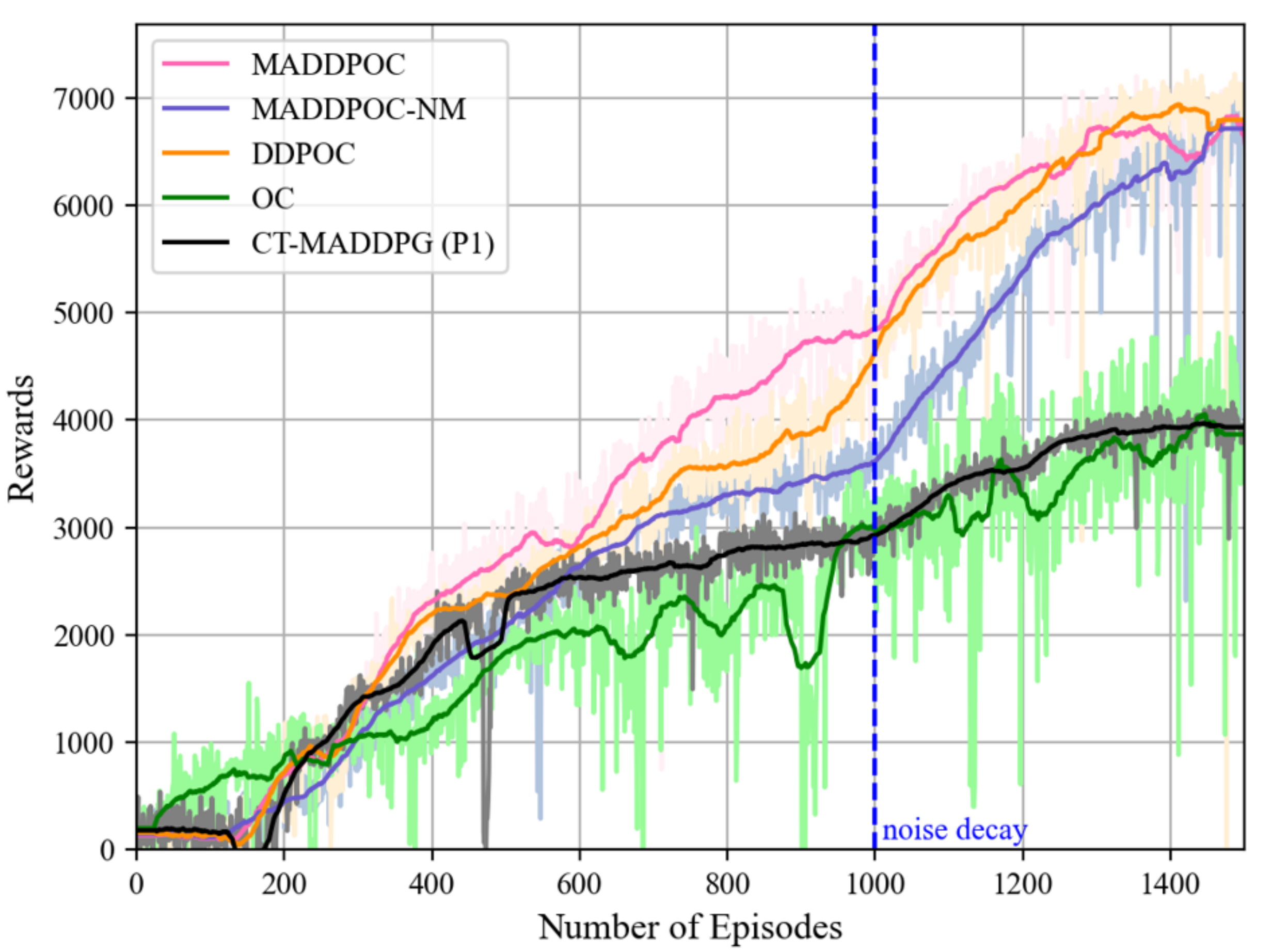}
	\caption{Rewards convergence in the training phase.}
	\label{fig:reward_convergence}
\end{figure}

Fig.~\ref{fig:reward_convergence} presents the rewards convergence for different algorithms in the training phase. Before noise decay, the MADDPOC's performance is the best, while DDPOC suffers from high action dimensions and converge a little bit slower. Besides, MADDPOC-NM is easily trapped into local optimal, i.e. prematurely terminate with surplus MUAV's energy. The reason lies in the hierarchical algorithm's inherent drawback of the mutual interference between upper and lower levels of decision-making. Specifically in our scenario, the over-option policy $\pi_{\Omega}$ cannot find appropriate charging points when the intra-option policy $\pi_{1,u}$ is not convergent, since the the charging process has a extremely demanding requirement for distance. To avoid violating the constraint (\ref{constaint2:energy}), the option of termination is easily selected, and thus, $\pi_{1,u}$ is unable to learn the later decisions. After noise decay, the accumulative rewards of all algorithms except OC have remarkable growth, since the same side constraints (\ref{constaint1:same_side}),  (\ref{constaint2:same_side}) are easily violated before. Note that the action deviations in OC architecture serve as trainable variables and artificial attenuation is impracticable,  which explains its poor performance in the noise-sensitive scenario. Meanwhile in this stage, MADDPOC-NM steps out of learning stagnation and finds other appropriate charging points. Moreover, CT-MADDPG cannot utilize the extra energy of MUAV and its throughput performance depends on the AUAV's energy.

\begin{figure}
	\centering
	\includegraphics[height=.8\columnwidth, width=1\columnwidth] {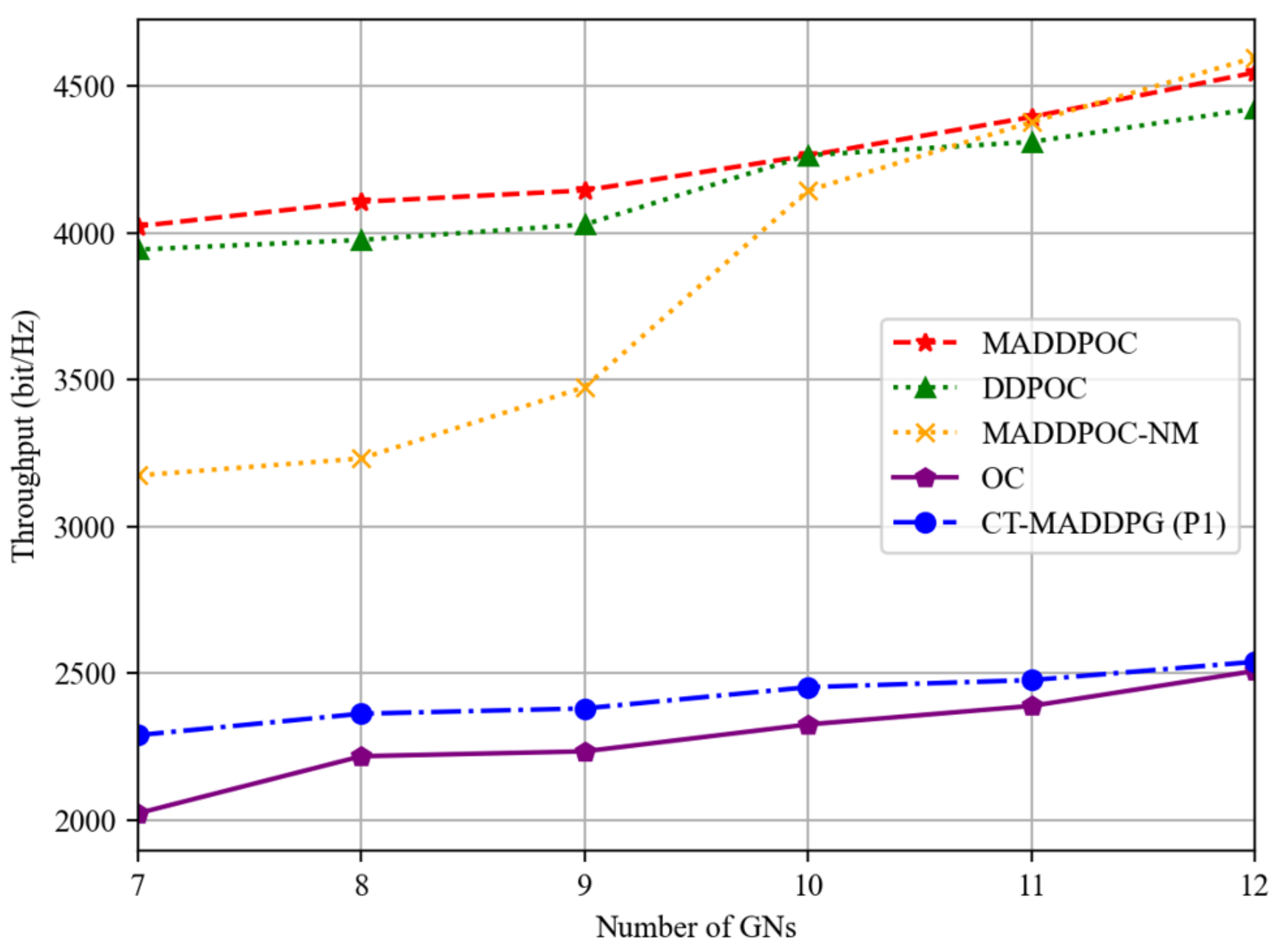}
	\caption{Accumulated throughput vs number of GNs.}
	\label{fig:throughput_vs_GN}
\end{figure}
In Fig.~\ref{fig:throughput_vs_GN}, we compare the total throughput obtained by aforementioned methods varying with different numbers of
GNs in deployment phase, where $\sigma_{\mathcal{A}} = 0$. We observe that MADDPOC has similar performance to
DDPOC. Meanwhile, the total throughput has an upward
trend, it is because that UAVs have higher probability of
covering users when the number of GN increases. Besides, the performance of MADDPOC-NM is not stable, since it may not step out of local optimal, and unable to take full advantage of the charging mechanism. Moreover, the OC's performance is even below CT-MADDPG's. Because OC could not make reasonable choices in our scenario and degenerate into PG, which suffers from high dimensional actions and the constraint (\ref{constaint2:same_side}).

\begin{figure}[t!]
	\centering
	\includegraphics[height=.8\columnwidth, width=1\columnwidth] {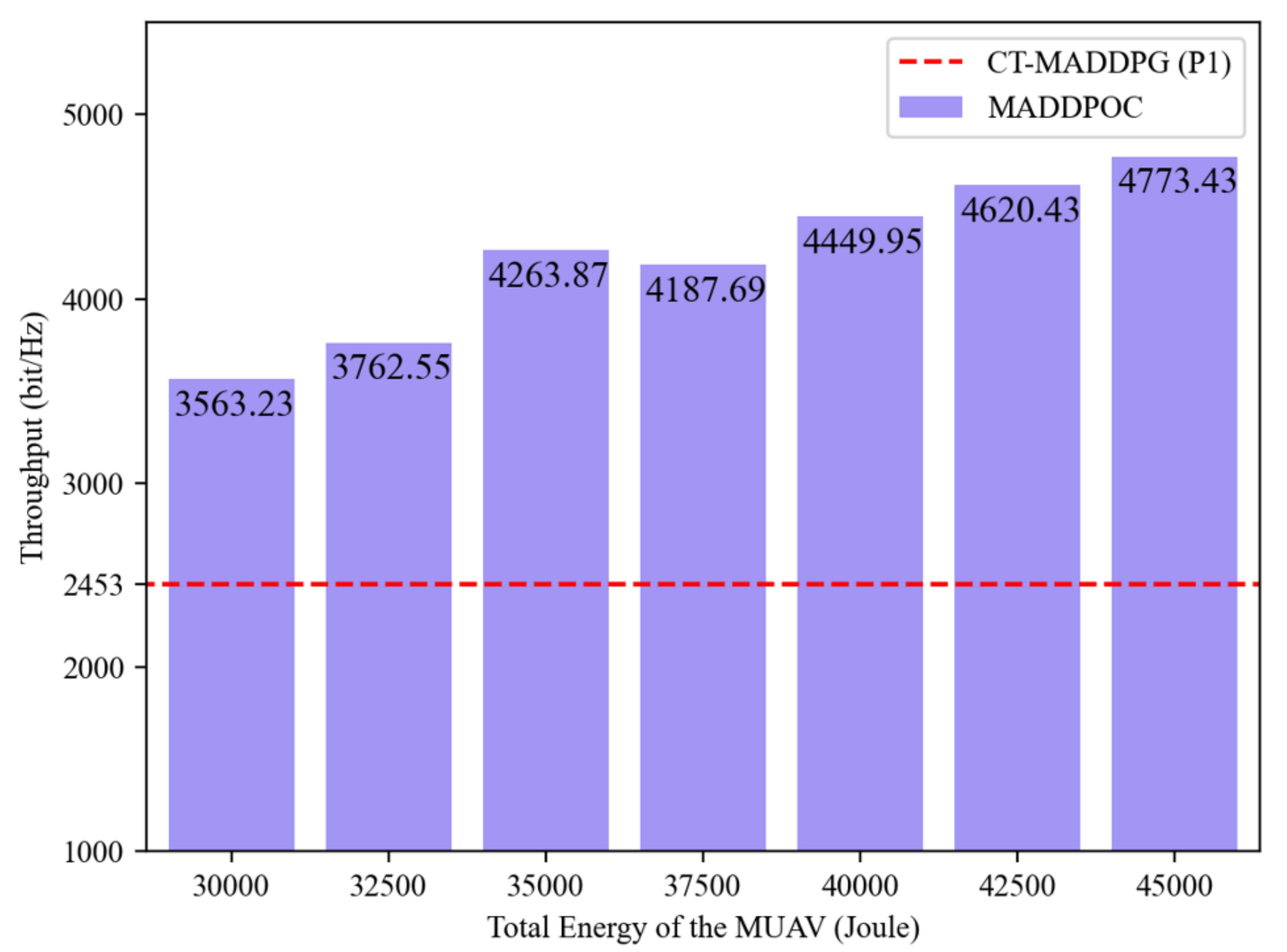}
	\caption{Accumulated throughput vs MUAV's max energy.}
	\label{fig:throughput_vs_EM}
\end{figure}
\begin{figure}
	\centering
	\includegraphics[height=.8\columnwidth, width=1\columnwidth] {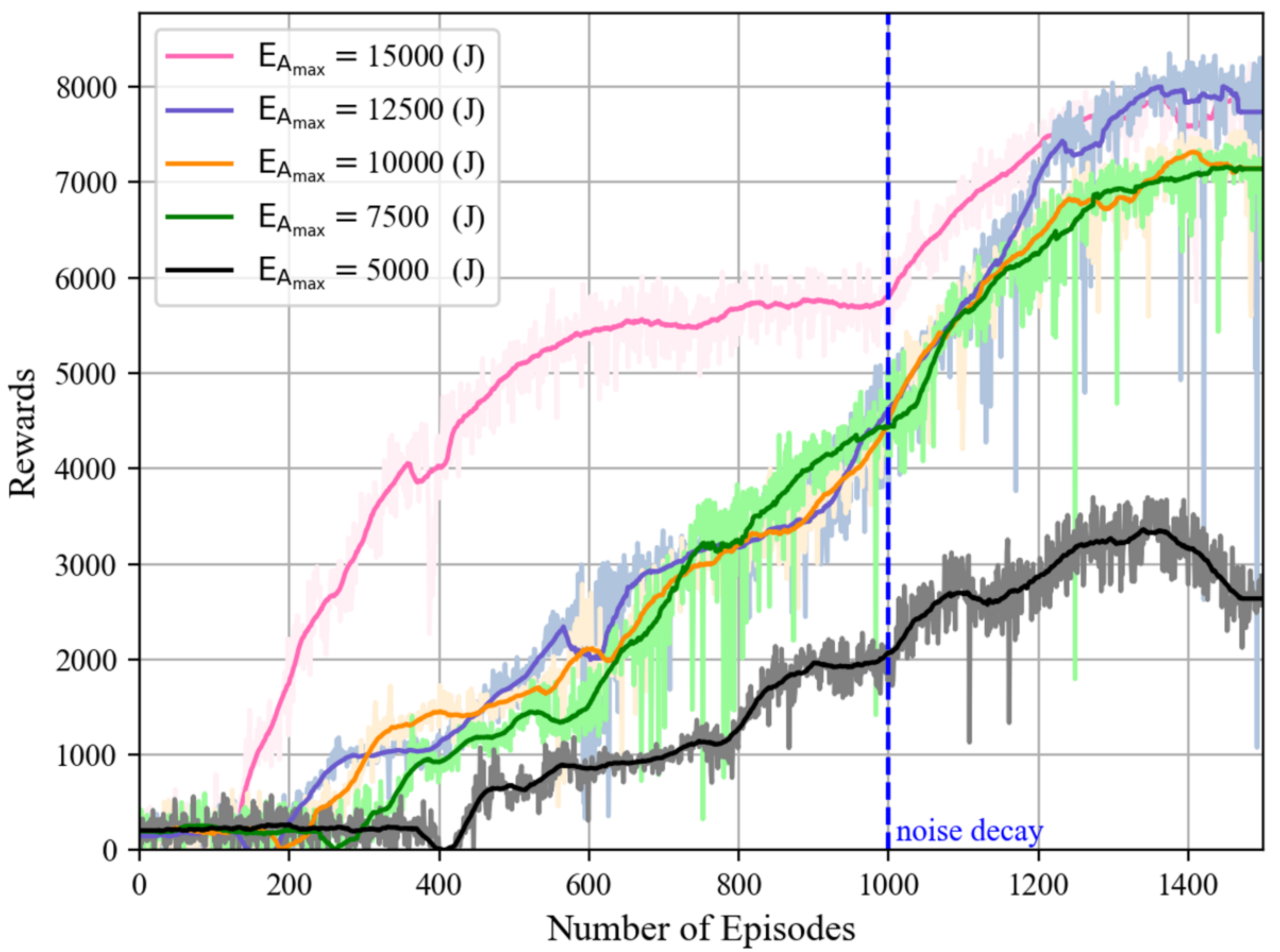}
	\caption{Reward curves for different AUAV's max energy.}
	\label{fig:reward_vs_EA}
\end{figure}
\begin{figure*}[]
	\subfigure[$E_{\rm{M}_{\rm{max}}}=35000$ Joules,  $E_{\rm{A}_{\rm{max}}}=15000$ Joules]
	{	
		\begin{minipage}{0.5\linewidth}
			\centering
			\includegraphics[height=0.8\columnwidth, width=1\columnwidth] {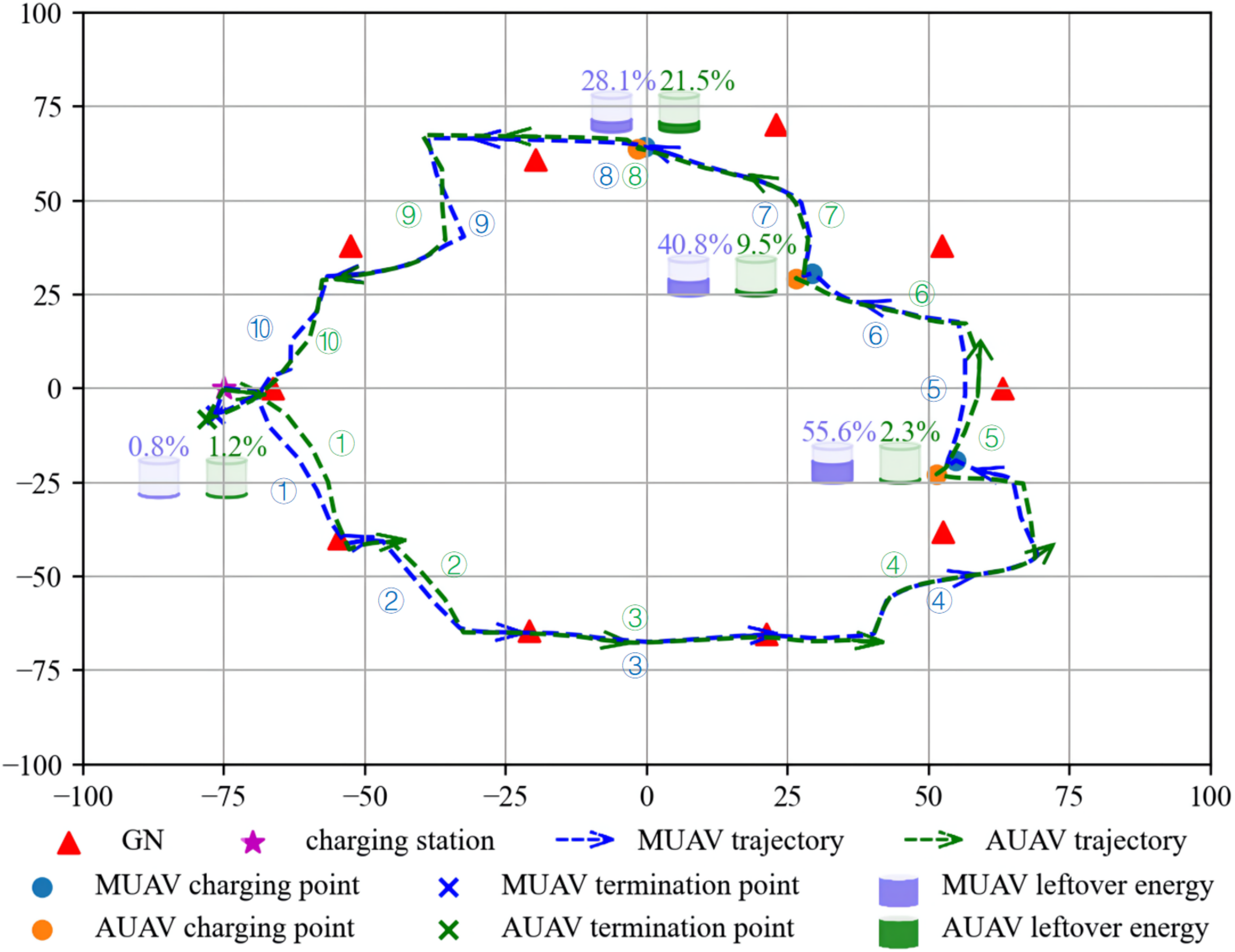}
			\label{fig:35000_15000_10} 
		\end{minipage}
	}
	\subfigure[$E_{\rm{M}_{\rm{max}}}=45000$ Joules,  $E_{\rm{A}_{\rm{max}}}=7500$ Joules]
	{	\begin{minipage}{0.5\linewidth}
			\centering
			\includegraphics[height=0.8\columnwidth, width=1\columnwidth] {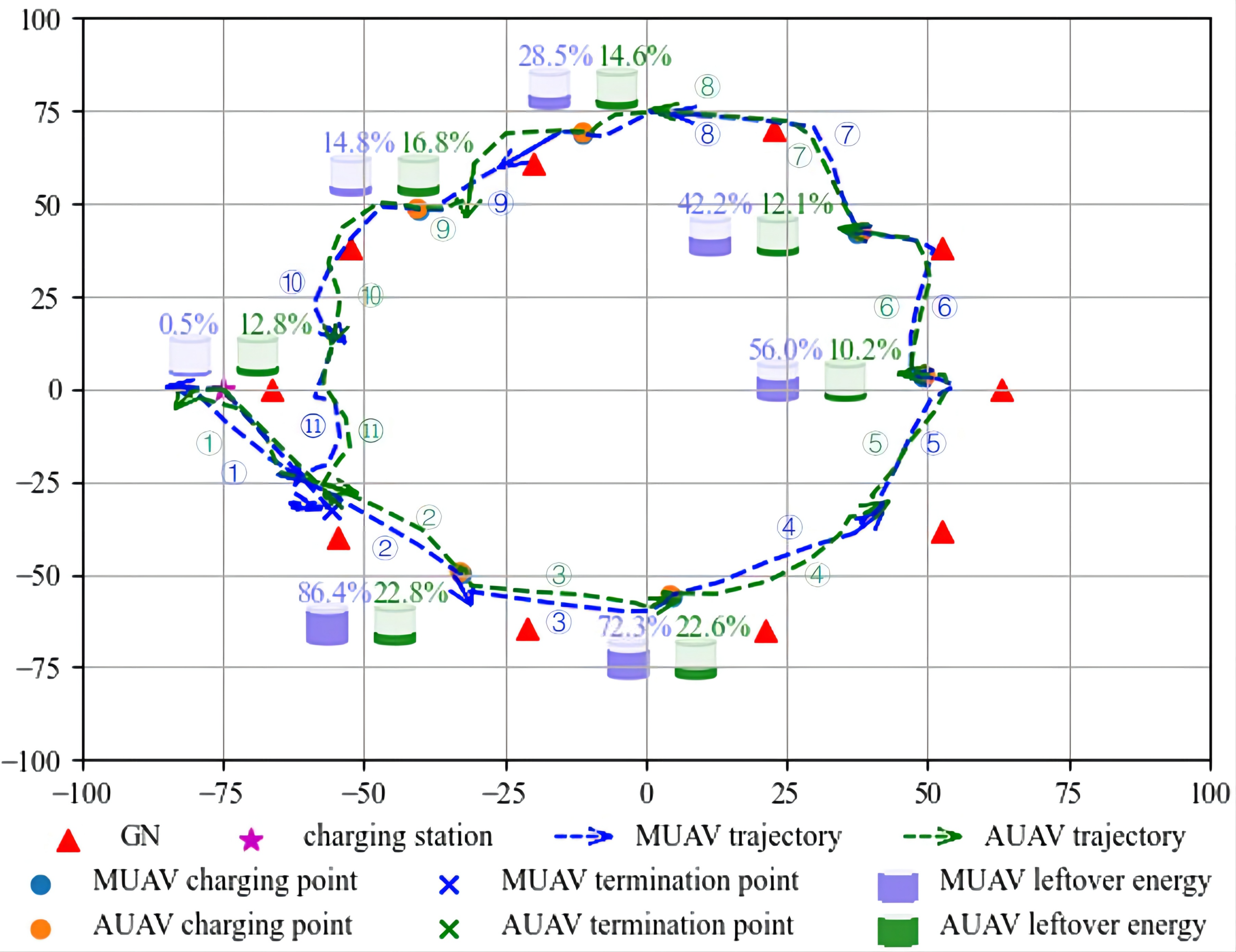}
			\label{fig:45000_7500_10} 
		\end{minipage}
	}
	\caption{The trajectories of charging scenario under different energy constraints.}
	\label{fig:trajectories_charging_scenario}
\end{figure*}

Fig.~\ref{fig:throughput_vs_EM} shows the impact of different MUAV's fully charged energy on the throughput performance. An interesting thing we observe is that the throughput performance does not always show an increasing trend as the max energy increases. There is even a decline when the max energy increases from 35000 Joules to 37500 Joules, of which the reason is twofold. On the one hand, it can be observed from Fig~\ref{fig:35000_15000_10} that both UAVs almost run out of energy simultaneously and terminate near the charging station in the former case. Agents tend to use similar trajectory policies if only the energy constraint changes. Thus in the latter case, the MUAV has about  $2500$ Joules left, close to the energy required for a single charging, which cannot be utilized. Because it's unable to keep the MUAV aloft without the energy. On the other hand, it's difficult for the over-option policy to determine whether to charge or terminate  when the AUAV's battery is low and the MUAV has just enough energy for a single charging, leading to the poor performance of the intra-option policy in the last few steps.

Similar situations can be seen in Fig.~\ref{fig:reward_vs_EA}, where the rewards convergence of MADDPOC varying with different AUAV's energy limits is shown and $E_{\rm{M}_{\rm{max}}}$ is set to $45000$ Joules. The throughput increases stepwise as the AUAV's max energy increases, since the energy-underutilized phenomenon also exists in the cases of $E_{\rm{A}_{\rm{max}}}=15000$ Joules and  $E_{\rm{A}_{\rm{max}}}=10000$ Joules. Another interesting thing is that the pink curve reaches a plateau at about 600 epochs, which is much faster than the others. Because the higher the AUAV's max energy, the lower the probability of constraint (\ref{constaint2:energy}) violation. Then more positive samples will be provided within an epoch, making the training process more efficient. Besides,  MADDPOC doesn't work when the AUAV's battery capacity becomes severe, e.g. $E_{\rm{A}_{\rm{max}}}=5000$ Joules, since the high-frequency violation of the constraint ({\ref{constaint2:energy}}) makes the training process extremely difficult.

\begin{figure}
	\centering
	\includegraphics[height=0.8\columnwidth, width=1\columnwidth] {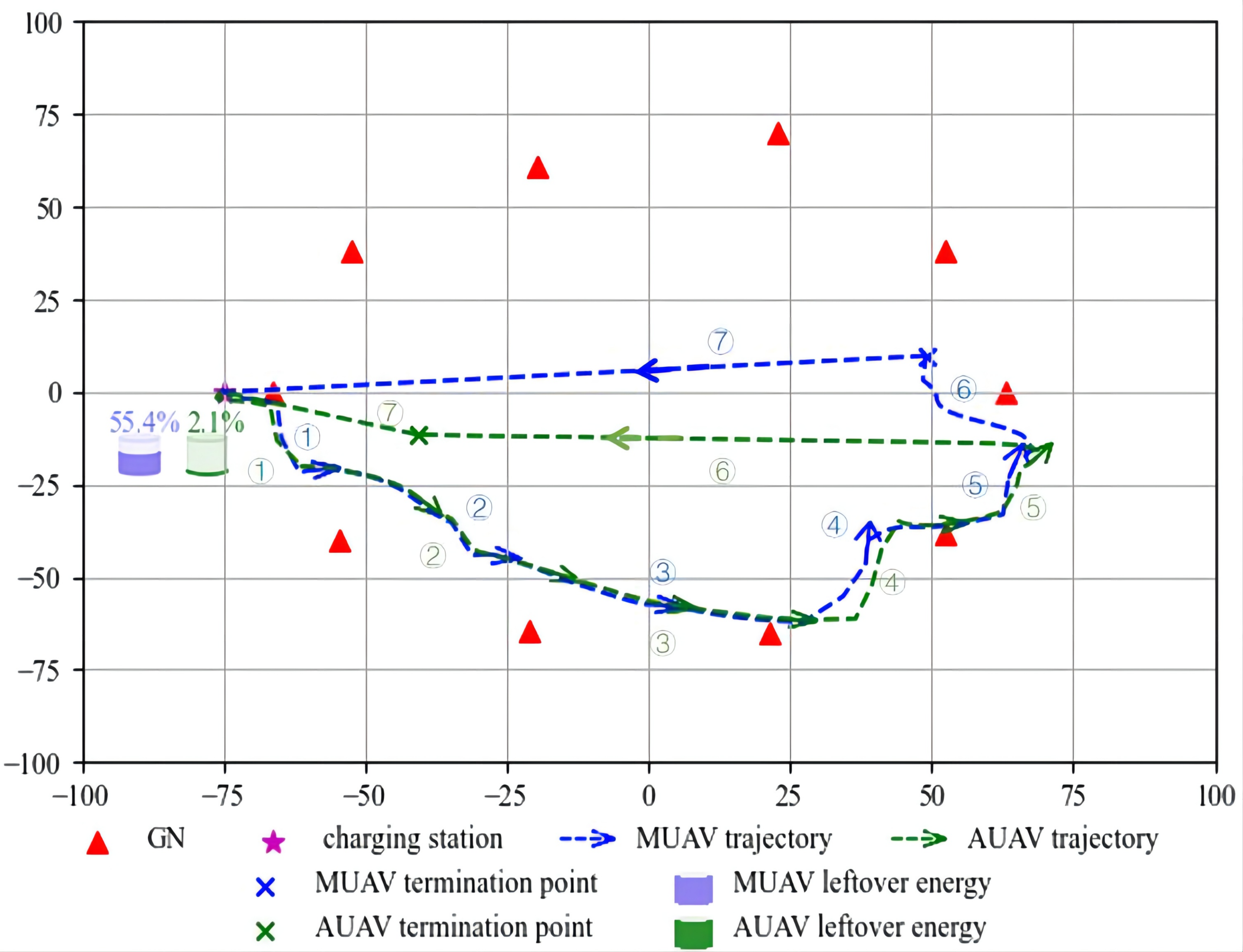}
	\caption{The trajectory of non-charging scenario.}
	\label{Trajectory:P1}
\end{figure}
Fig.~\ref{Trajectory:P1} plots the trajectory of non-charging scenario. The AUAV can only serve half of GNs and have to turn back midway. We observe that the AUAV will terminate somewhere close to the charging station, while the MUAV keeps quasi-stationary over the linked GN until its termination. On the one hand,  there is a trade-off between termination and subsidiary communication for the AUAV, while the MUAV with high battery capacity only needs to focus on providing high quality service. On the other hand, the range of MUAV's operation area is strictly restrained by (\ref{constaint1:coverage_range}), so it cannot leave too far from the linked GN.

Fig.~\ref{fig:trajectories_charging_scenario} shows the  trajectories of charging scenario under different energy constraints. Note that the segment number pair, e.g. the \ding{172} in blue and green, represents the flight trajectories of two UAVs in the same period of time. The leftover energy of UAVs at each charging point as well as after termination is also labeled in terms of cylindrical icon. It can be observed that two trajectories almost coincide anytime, which depends on a high-degree cooperation between two actors $\pi_{1, \rm{M}}$ and $\pi_{1, \rm{A}}$. Because the data rate can be maximized when the IRS is close enough to the transmitter or the receiver. In this case, the constraint (\ref{constaint1:same_side}) is easy to be violated even with a low action deviation, which  supports the observation of Fig.~\ref{fig:reward_convergence}. Besides, the leftover energy of both UAVs remains positive during the operation time, which guarantees the validity of
trajectories. We also observe that MADDPOC always ends up with one or both ultimate energy of two UAVs since it's easy to judge from the positivity and negativity of the state information $E_{\mathrm{M}l}[n]$ and $E_{\mathrm{A}l}[n]$.

The numerical results under different energy constraints, including the items of total operation slots, charging count, leftover energy, accumulative throughput and energy efficiency are summarized in Table~\ref{tab:numerical_result}. It can be observed that MADDPOC is unable to take full advantage of both UAVs' energy in most cases. Especially in the 4th case, working hours barely changes whether charge three or four times. Because it no longer makes sense to charge when the MUAV's leftover energy is just above the threshold for a single charging $p_c\delta{\tau}$ and the AUAV is running out of energy. The other key factor affecting energy efficiency is the termination location. The farther away UAVs are from the charging station at the termination step, the more energy UAVs wastes to return.

\begin{table*}[]
	\centering
	\caption{Numerical results  under different energy constraints.}
	\setlength{\tabcolsep}{2mm}{
		\begin{tabular}{|c|c|c|c|c|c|c|c|c|}
			\hline
			Index&$E_{\rm{M}_{max}}$ (J)  &$E_{\rm{A}_{max}}$ (J) &$N_\tau$ & Charging count & $E_{\rm{M}}\left[ N_\tau\right] $ ($\%$) & $E_{\rm{A}}\left[ N_\tau\right]$ ($\%$) & Throughput (bit/Hz)&Energy efficiency (bit/Hz/Joule)\\  \hline
			1&30000 &\multirow{8}{*}{15000}  &270                       &2                       &5.05$\%$                       &$<1\%$                       &3563     &0.0792                \\  \cline{1-2}\cline{4-9}

			2&32500 & & 291                       &3                       &$<1\%$                       &15.03$\%$                       & 3762        &0.0792           \\ \cline{1-2}\cline{4-9}

			3&35000 & &316                       &3                       &$<1\%$                       &$1.2\%$                       &4263         &\textcolor[rgb]{1.00,0.00,0.00}{0.0853}           \\ \cline{1-2}\cline{4-9}
			
			\multirow{2}{*}{4} & \multirow{2}{*}{37500} 
			& &317                       &3                       &7.39$\%$                      &1.40$\%$                       &4187          &0.0798            \\ 
			\cline{4-9}
			& & &318                       & 4                     &$<1\%$                       &17.37$\%$                       &4151          &0.0791             \\ \cline{1-2}\cline{4-9}
			
			5&40000 & &340                       &4                      &$<1\%$                       &8.57$\%$                       &4449          &0.0809             \\ \cline{1-2}\cline{4-9}
			
			6&42500 & &358                       &4                      &3.67$\%$                       &2.31$\%$                       &4620        &0.0803            \\ \cline{1-2}\cline{4-9}
			
			7&\multirow{5}{*}{45000}  & &368                      &5                       &1.25$\%$                       &13.53$\%$                       &4773          &0.0796             \\ \cline{1-1}\cline{3-9}
			
			8& &12500 &365                      &5                      &$<1\%$                       &$<1\%$                      &4725           &0.0822             \\ \cline{1-1}\cline{3-9}
			
			9& &10000 &343                       &5                      &$5.80\%$                       &$<1\%$                      &4439          & 0.0807            \\ \cline{1-1}\cline{3-9}
			
			10 & &7500 &338                       &6                      &$<1\%$                       &$12.76\%$                      &4389          &\textcolor[rgb]{1.00,0.00,1.00}{0.0836}             \\ \cline{1-1}\cline{3-9}
			
			11 & &5000 &187                      &10-13                   &$\backslash$                     &$\backslash$                      &2299          & 0.0460            \\ \hline

	\end{tabular}}
	\label{tab:numerical_result}
\end{table*}

\section{Conclusions and Future work}\label{conclusion}
In this paper, we have investigated a MUAV-powered IoT network, in which we have proposed using a rechargeable AUAV equipped with an IRS to enhance the communication signals from the MUAV and also leverage the MUAV as a recharging power source. Under the proposed model, we have investigated the optimal collaboration strategy of these  energy-limited UAVs to maximize the accumulated throughput of the IoT network. Depending on whether there is charging between the two UAVs, two optimization problems have been formulated. To solve them, we have proposed the CT-MADDPG algorithm to enable the UAVs' cooperative communication and the MADDPOC algorithm, which supports the multi-agent hierarchical control. The numerical results show that MADDPOC performs better than the benchmarks in terms of the convergence rate and the accumulative throughput. Furthermore, we also observed that the throughput performance does not always show an increasing trend as the fully charged energy increases because of the existence of energy-underutilization phenomenon.

Although we deploy an IRS-equipped AUAV to help with communications, the phase optimization of IRS is replaced by MRT for simplicity. Given the dimensional differences among actions, it's not recommended to output the trajectories,  transmit power and the phase of IRS simultaneously with a single network, since the phase with excessive dimension severely affects the performance of the other two low-dimensional actions. Thus, another phase-dedicated network is needed to be pre-trained before the optimization of trajectories and transmit power. In the case of perfect channel prediction, the channel matrix is formulated by the positions of UAV, IRS and linked GN, which can be effectively mapped to the optimized phase by deep transfer learning \cite{9367008}, while in the unperfect case, additional information from past channels is needed \cite{9110869}. For the future work, we will investigate the trajectory design
and phase optimization of multiple IRS-equipped AUAVs in a MUAV-powered IoT network.

\appendices
\section{Optimal Phase Shift Control at the IRS}\label{appendix}
In the appendix, we derive the optimal phase control strategy at the IRS in our scenario. In the case of MRT, the signal noise ratio in (\ref{eq:data_rate})  can be formulated as:
\begin{align}                                            \gamma_{s}[n]&=\left\|\left(\mathbf{h}_\mathrm{MG_k}\left [ n \right ] \right)^{H}\mathbf{w}_{k}\left [ n \right ]\right\|^2  /\sigma^2\nonumber \\ &=\frac{S_{\mathrm{R}}^2S_{\mathrm{M}}}{\sigma^2}\left\|\left(\mathbf{h}_{\mathrm{RG}_k}\left [ n \right ] \right)^{H}\mathbf{\Phi}\left[n \right]\mathbf{h}_\mathrm{RM}\left [ n \right ]\right\|^2\nonumber \\ &=\frac{S_{\mathrm{R}}^2S_{\mathrm{M}}}{\sigma^2}\left\|\sum_{x=0}^{S_{{\rm{R}}_x}}\sum_{y=0}^{S_{{\rm{R}}_y}}e^{j(\varphi^{\mathrm{RG}}_{xy}[n]+\varphi_{x,y}[n]+\varphi^{\mathrm{RM}}_{xy}[n])}\right\|^2.\label{eq:reduced_channel}
\end{align}
Note that
$\mathbf{h}_\mathrm{RM}\left [ n \right ]=\mathbf{a}\left(\theta_{\rm{MR}}^{\rm{A}}\left [ n \right ], \xi_{\rm{MR}}^{\rm{A}}\left [ n \right ], S_{{\rm{R}}_x}, S_{{\rm{R}}_y} \right)$ according to (\ref{eq:H_MR}), $\varphi^{\mathrm{RG}}_{xy}[n], \varphi^{\mathrm{RM}}_{xy}[n]$ denote the phases of $xy$-th elements of vectors $\left( \mathbf{h}_{\mathrm{RG}_k}\left [ n \right ]\right)  ^{H}, \mathbf{h}_\mathrm{RM}\left [ n \right ]$, respectively, and $\varphi_{x,y}[n]$ donates the phase of IRS's $x,y$-th unit. Let $\varphi_{xy}^\mathrm{sum}[n]$ represent the sum of the three phases. The formula (\ref{eq:reduced_channel}) satisfies the Cauchy inequality:
\begin{align} 
	\left\|\sum_{x=0}^{S_{{\rm{R}}_x}}\sum_{y=0}^{S_{{\rm{R}}_y}}e^{j\varphi_{xy}^\mathrm{sum}[n]}\right\|\leq \sum_{x=0}^{S_{{\rm{R}}_x}}\sum_{y=0}^{S_{{\rm{R}}_y}}\left\|e^{j\varphi_{xy}^\mathrm{sum}[n]}\right\|,
\end{align}
equal if and only if $\varphi_{xy}^\mathrm{sum}[n]=\varphi, \forall x, y, n$, where $\varphi\in[0,2\pi)$ is a constant. Let $\varphi=0$, we can get $\varphi_{x,y}[n]=-(\varphi^{\mathrm{RG}}_{xy}[n]+ \varphi^{\mathrm{RM}}_{xy}[n])$. Then the optimal phase in (\ref{eq:IRS_phase}) can be obtained by substituting the specific expressions in (\ref{eq:H_MR}) and (\ref{eq:h_RGk}).

\bibliography{IEEEabrv,jingren_journal}
\bibliographystyle{IEEEtran}
\end{document}